\documentclass[referee]{sn-jnl}

\usepackage{graphicx}%
\usepackage{multirow}%
\usepackage{amsmath,amssymb,amsfonts}%
\usepackage{amsthm}%
\usepackage{mathrsfs}%
\usepackage[title]{appendix}%
\usepackage{xcolor}%
\usepackage{textcomp}%
\usepackage{manyfoot}%
\usepackage{booktabs}%
\usepackage{algorithm}%
\usepackage{algorithmicx}%
\usepackage{algpseudocode}%
\usepackage{listings}%
\usepackage{bm}%
\begin{document}

\title{DiffCrysGen: A Generative Diffusion Model for Accelerated Design of Inorganic Crystalline Materials}

\author[1,2]{\fnm{Sourav} \sur{Mal}}

\author[3]{\fnm{Nehad} \sur{Ahmed}}

\author[1,2]{\fnm{Junaid} \sur{Jami}}

\author[4,2]{\fnm{Subhankar} \sur{Mishra}}\email{smishra@niser.ac.in}

\author[1,2]{\fnm{Prasenjit} \sur{Sen}}\email{prasen@hri.res.in}

\affil[1]{\orgname{Harish-Chandra Research Institute},
\orgaddress{\street{Chhatnag Road, Jhunsi},
\city{Prayagraj},
\postcode{211019},
\country{India}}}

\affil[2]{\orgname{Homi Bhabha National Institute},
\orgaddress{\street{Training School Complex, Anushakti Nagar},
\city{Mumbai},
\postcode{400094},
\country{India}}}

\affil[3]{\orgdiv{Department of Chemistry},
\orgname{Indian Institute of Science Education and Research (IISER) Tirupati},
\orgaddress{\city{Tirupati},
\state{Andhra Pradesh},
\postcode{517619},
\country{India}}}

\affil[4]{\orgdiv{School of Computer Sciences},
\orgname{National Institute of Science Education and Research (NISER)},
\orgaddress{\city{Jatni},
\state{Odisha},
\postcode{752050},
\country{India}}}

\abstract{Efficient exploration of the vast chemical space is a fundamental challenge in materials design and discovery, particularly for designing functional inorganic crystalline materials with targeted properties. Diffusion-based generative models have emerged as a powerful route, but most existing approaches require domain-specific constraints and separate diffusion processes for atom types, atomic positions, and lattice parameters, adding complexity and limiting efficiency. Here, we present DiffCrysGen, a fully data-driven, score-based diffusion model that generates complete crystal structures in a single, end-to-end diffusion process. This unified framework simplifies the model architecture and accelerates sampling by two to three orders of magnitude compared to existing methods without compromising chemical and structural diversity of the generated materials. In order to demonstrate the efficacy of DiffCrysGen in generating valid and useful materials, using density functional theory (DFT), we validate a number of newly generated rare earth-free magnetic materials that are energetically and dynamically stable, and are potentially synthesizable. These include ferromagnets with high saturation magnetization and large magnetocrystalline anisotropy, as also metallic antiferromagnets. These results establish DiffCrysGen as a general platform for accelerated design of functional materials.}


\maketitle

\section*{Introduction}

The ability to generate entirely new, stable materials without human bias could revolutionize fields like energy storage, catalysis, and quantum materials. Designing such high-performance materials with targeted properties is challenging due to the astronomically large chemical and structural design space. Traditional materials design relies on trial-and-error experimentation and human intuition, typically modifying known compounds, such as via elemental substitutions or introducing site-specific dopants. Properties of these candidate materials are then evaluated via high-throughput density functional theory (DFT) calculations. This process is ultimately constrained by the high computational cost of DFT.

Availability of large, high-quality materials databases in recent years~\cite{MP,OQMD,JARVIS,aflow,Alex-1,Alex-2,Alex-3} has enabled machine learning (ML) to accelerate the 
design by predicting material properties with high accuracy~\cite{CGCNN,MEGNet,ALIGNN}, thereby reducing the number of candidates requiring expensive DFT validation. However, most workflows still construct initial candidates by modifying known materials, a strategy lacking universal design rules, making them dependent on prior human knowledge and limiting scalability and innovation. Generative ML models offer a path to overcome this limitation by directly learning the distribution of crystal structures and enabling the {\it de novo} design of materials without predefined templates.

Early generative models for crystals, principally those employing variational autoencoders (VAEs)~\cite{VAE1,VAE2,iMatGen,FTCP,CDVAE,ACS-JCIM,Court-ChemMat} and generative adversarial networks (GANs)~\cite{GAN,DCGAN,CubicGAN,ZeoGAN,CrystalGAN,CCCG, PGCGM}, have demonstrated considerable potential, enabling the design of novel materials across a range of applications~\cite{CSFE,superconductor,Lyngby2022,MagGen}.
However, these approaches often suffer from drawbacks, including 
unstable training and limited diversity in generated structures.

Diffusion models~\cite{sohldickstein,DDPM,song,EDM}, which iteratively denoise random noise into realistic samples, have recently surpassed these methods in generating diverse and high-quality crystal structures.
However, state-of-the-art crystal diffusion models~\cite{mattergen,DiffCSP,DiffCSP++,chemeleon} often require separate diffusion processes for atom types, atomic positions, and lattice parameters, each with its own latent space. Structural correlations are then introduced only during reverse denoising process via conditional coupling, which complicates the overall framework and slows down the sampling for new crystals generation. 

Here, we present DiffCrysGen, a fully data-driven, score-based diffusion model that unifies the generation of all structural components into a single end-to-end diffusion process. The unified latent space of the model naturally captures inter-dependencies among the different components, reducing the computational overhead and simplifying the framework. We demonstrate its ability in generating chemically valid, diverse inorganic crystalline materials at sampling speeds two to three orders of magnitude faster than the existing models.

As a proof of concept, we target the urgent challenge of designing rare earth (RE)-free magnets, critical for sustainable energy technologies but constrained by natural and geopolitical factors. Structures generated by DiffCrysGen were screened using ML property predictors for thermodynamic stability and magnetization, followed by rigorous DFT validation on selected candidates, including structural relaxation, magnetic property evaluation, and phonon stability analysis. This led to the identification of multiple previously unreported RE-free magnetic materials which include both ferromagnets with high saturation magnetization and magnetocrystalline anisotropy, and metallic
antiferromagnets.

Our work bridges generative AI, machine learning and inorganic materials design, introducing a generalizable platform for the {\it de novo} design of functional materials. By performing structure generation within a unified diffusion-based framework, DiffCrysGen accelerates the exploration of crystal chemical space and enables the design of novel materials with important implications for all areas of technology requiring advanced materials.

\section*{Results}

\subsection*{Diffusion model for crystalline materials}

DiffCrysGen is a score-based diffusion model developed within the stochastic differential equation (SDE) framework~\cite{song}. It is trained on a compact 2D point cloud representation that jointly encodes atom types, lattice parameters, and atomic coordinates, as detailed in the Methods section. The model is trained on a large curated dataset of 151,294 optimized crystal structures from the Alexandria database~\cite{Alex-1,Alex-2,Alex-3}, each containing up to 20 atoms per unit cell. The size of this dataset is crucial, enabling our model to implicitly learn complex crystallographic priors directly from data without reliance on handcrafted constraints.

\begin{figure}
    \centering
    \includegraphics[scale=0.25]{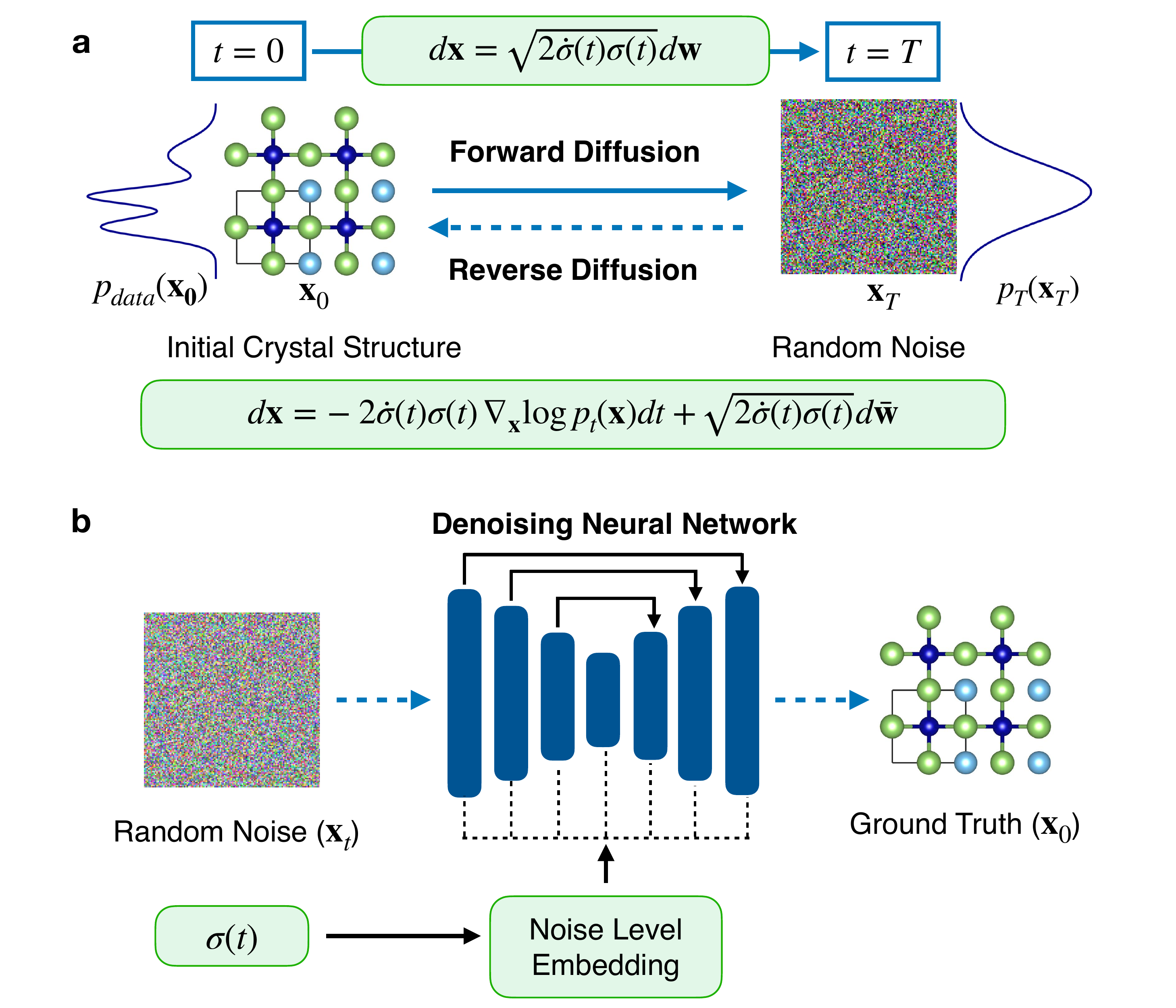}
    \caption{{\bf Generative diffusion framework in DiffCrysGen.} {\bf a} The diffusion process consists of two stages: forward diffusion (corruption) and reverse diffusion (denoising), both governed by the corresponding stochastic differential equations (SDEs) with a time-dependent noise level $\sigma(t)$. Here, $p_{data}({\bf x}_0)$ and $p_{T}({\bf x}_T)$ denote initial data distribution and prior distribution, respectively. The reverse diffusion is guided by the score function $\nabla_{\bm x} \log p_t({\bm x})$ of the marginal probability density $p_t(\bm x)$ at each time step $t$. {\bf b} Schematic of the denoising process implemented by a noise-conditional UNet as the denoising neural network. It takes as input a noisy data ($\bm{x_t}$) and corresponding noise level ($\sigma_t$), and predicts the ground truth denoised structure ($\bm{x_0}$).         
    }
    \label{diffusion-schematic}
\end{figure}

The diffusion process involves two stages: forward and reverse diffusion. In the forward diffusion, Gaussian noise is progressively added to corrupt the input structure, transforming the complex data distribution into a simple Gaussian prior. The reverse diffusion process is the generative phase, where noise is gradually removed from an initial random sample to reconstruct a valid crystal structure. This process is illustrated schematically in Fig.~\ref{diffusion-schematic}a. To enable reverse diffusion, we train a denoising neural network, based on a noise conditional UNet that takes as input the noisy structure and its noise level, and predicts the underlying clean data, schematically shown in Fig.~\ref{diffusion-schematic}b. The underlying theory and implementation details of DiffCrysGen are provided in the Methods section.

Key characteristics of DiffCrysGen are its conceptual simplicity and ease of implementation: it treats the entire materials representation with a single, unified diffusion process, allowing a single denoising neural network to predict a holistic score for the entire noisy crystal data. This is unlike graph-based models which treat atomic-types, fractional coordinates and lattice vectors through separate diffusion processes.
This unified treatment not only significantly simplifies DiffCrysGen's architecture but also dramatically improves its computational efficiency. Quantitatively, DiffCrysGen comprises only 1.3 million trainable parameters, making it an order of magnitude (or more) smaller than other recent diffusion-based generative models (see Table~\ref{tab:model_sizes} for a detailed comparison). This positions DiffCrysGen as one of the most lightweight diffusion-based generative models for crystalline materials reported to date. One practical benefit of this efficiency is its extremely fast generation speed: 100,000 crystal structures are generated in only 5.41 minutes on a single NVIDIA H100 GPU. This corresponds to a generation rate of $308.07$ samples/sec.

\begin{table}[h]
\centering
\caption{Comparative analysis of diffusion-based generative model sizes}
\vspace{0.2cm} 
\begin{tabular}{lcc}
\hline
\textbf{Model} & \textbf{Total Trainable Parameters} \\
\hline
\textbf{DiffCrysGen} & \textbf{1.3 M} \\ 
DiffCSP~\cite{DiffCSP}, DiffCSP++~\cite{DiffCSP++} & 12.3 M  \\
MatterGen~\cite{mattergen} & 46.8 M  \\
\hline
\end{tabular}
\label{tab:model_sizes}
\end{table}

\subsection*{Comparative Benchmarking Against State-of-the-Art Crystal Diffusion Models}

\begin{table}[t]
\centering
\caption{Definitions and interpretation of crystal generative model evaluation metrics for benchmarking.}
\label{tab:metrics}
\begin{tabular}{p{3.5cm} p{8.5cm} p{3cm}}
\toprule
\textbf{Term} & \textbf{Definition} & \textbf{Interpretation} \\
\midrule

Validity Rate (V) 
& Percentage of generated structures with a minimum pairwise interatomic distance greater than 0.5~\AA. 
& Higher is better ($\uparrow$) \\

Negative $h_{\mathrm{form}}$ Rate 
& Percentage of generated structures with $h_{\mathrm{form}} \leq 0$ eV/atom after optimization. 
& Higher is better ($\uparrow$) \\

Stability Rate (S) 
& Percentage of generated structures with $E_{\mathrm{hull}} \leq 0.1$ eV/atom after optimization. 
& Higher is better ($\uparrow$) \\

S.U.N. Rate 
& Percentage of generated structures that are stable ($E_{\mathrm{hull}} \leq 0.1$ eV/atom), unique, and novel. A structure is considered unique if no other generated structure shares the same composition and space group, and novel if it does not appear in the training set under the same criteria.
& Higher is better ($\uparrow$) \\

Average RMSD 
& Average root-mean-square displacement (RMSD) of atomic positions between generated structures and their corresponding optimized structures, expressed in \AA. 
& Lower is better ($\downarrow$) \\

P1 Rate (P) 
& Percentage of generated structures with P1 space group before optimization. 
& Lower is better ($\downarrow$) \\

Generation Rate (G) 
& Number of generated structures per second. 
& Higher is better ($\uparrow$) \\

Effective Generation Rate (EGR) 
& 
The rate at which a model generates S.U.N. structures per second.
& Higher is better ($\uparrow$) \\

\bottomrule
\end{tabular}
\label{tab:benchmark}
\end{table}

We start by benchmarking DiffCrysGen against MatterGen~\cite{mattergen} and DiffCSP~\cite{DiffCSP} using a comprehensive set of evaluation metrics, summarized in Table~\ref{tab:benchmark}. There are also a few recent symmetry-aware generative models such as DiffCSP++~\cite{DiffCSP++}, SymmCD~\cite{SymmCD}, WyckoffDiff~\cite{WyckoffDiff}, which can perform symmetry-constrained crystal generation through group-theoretic parameterizations and Wyckoff-position-based representations. Their primary objective is to enforce crystallographic symmetry directly within the generative process. DiffCrysGen, in contrast, is intentionally designed as a general-purpose, symmetry-agnostic diffusion framework. Rather than explicitly constraining space-group symmetry, our aim was to make the model learn structural regularities implicitly from data through a unified representation of atomic species, coordinates, and lattice parameters. 

The commonly used metrics in the literature - namely the validity rate (V), negative formation energy  rate, stability rate (S), Stable, unique, and novel (S.U.N.) rate, and average RMSD (definitions are given in Table~\ref{tab:benchmark}), are employed. In addition, we introduce three new evaluation metrics that overcome limitations of the current benchmarking practices, and provide a more rigorous assessment of generative models. 

P1 rate (P) quantifies the percentage of generated structures assigned to the P1 space group prior to structural optimization. Existing generative models are biased towards generating structures with P1 space group and it is a well-known limitation. By explicitly introducing the P1 rate, we enable direct assessment of a models’s ability to learn and generate crystallographic symmetry. Lower values of P are desirable. 

Generation rate (G) measures the number of structures generated per second and directly reflects the computational throughput of a generative model. To the best of our knowledge, this metric has not been reported or systematically compared in prior studies, despite its central importance for large-scale materials design and discovery. Incorporating G explicitly allows benchmarking to account for computational efficiency, with higher values indicating better throughput.

Effective generation rate (EGR) combines structural quality and computational efficiency into a single throughput-aware metric, defined as 
\begin{equation}
\mathrm{EGR} = \mathrm{G} \times \mathrm{S.U.N.} \quad (\text{structures/s})
\end{equation}
This metric represents the rate at which S.U.N. structures are produced, thereby providing a realistic measure of a model’s utility in practical design pipelines. We believe EGR captures an essential but previously overlooked aspect of generative model performance and offers a natural basis for a fair comparison across models with differing trade-offs between speed and accuracy. We hope that the introduction of these metrics will help establish more comprehensive and practically relevant benchmarking standards for future generative models in materials science.

Before proceeding further, we would like to emphasize that all calculations were performed under an 
identical computational protocol for all compared models to ensure a fair benchmark. We randomly selected 4,000 generated structures from each model and optimized them using the same MLFF workflow (details provided in the methods section), under identical computational settings on the same H100 GPU. We used a consistent batch size of 1,000 across all models to avoid bias arising from hardware utilization or batching differences. For each of these models, we  used the respective training set for novelty evaluation. The RMSD values are calculated between each generated structure and its optimized version after relaxation. The $E_{hull}$  values are computed with respect to the same set of reference stable phases from the Materials Project~\cite{MP} database for all compared models.
This standardized setup ensures that the reported comparisons reflect intrinsic model characteristics rather than discrepancies in computational configuration.

\begin{figure}
    \centering
    \includegraphics[scale=0.6]{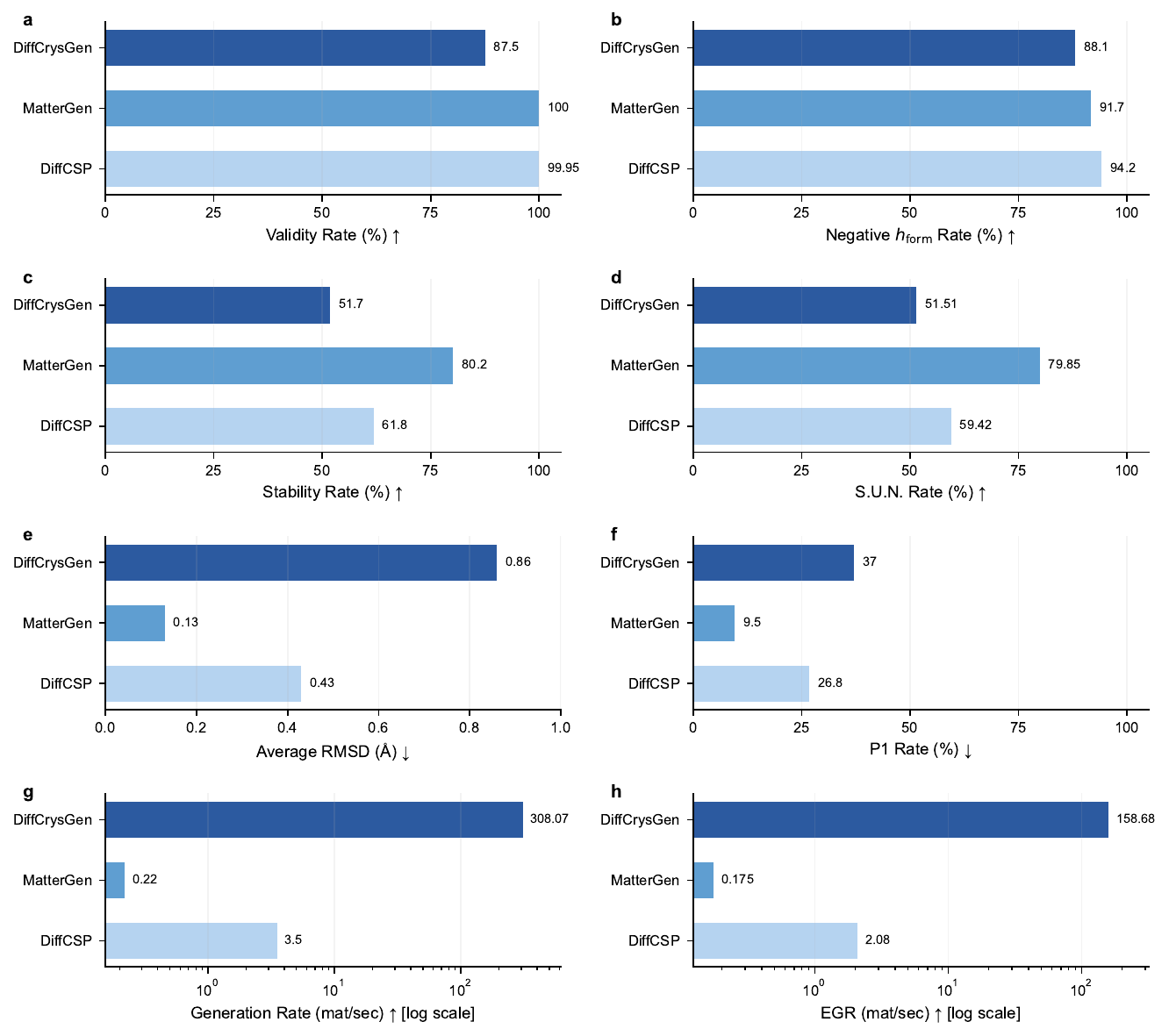}
    \caption{{\bf Benchmark comparison of DiffCrysGen with other diffusion-based generative models.} Benchmarking is performed using multiple evaluation metrics defined in Table~\ref{tab:benchmark}.: (a) validity rate, (b) negative $h_{form}$ rate, (c) stability rate, (d) stable, unique, and novel (S.U.N.) rate, (e) average RMSD, (f) P1 rate, (g) generation rate, and (h) effective generation rate (EGR). Upward (downward) arrows indicate that higher (lower) values correspond to better performance for the respective metric.}
    \label{benchmark_plot}
\end{figure}

Fig~\ref{benchmark_plot} presents a comparison of the models on a number of metrics.  
DiffCrysGen is competitive with MatterGen and DiffCSP on most of the stability metrics though it lags somewhat in RMSD. 
On a wider landscape, the RMSD values for DiffCrysGen are of the same order of magnitude as those for MatterGen and DiffCSP, 
while PG-SchNet, G-SchNet~\cite{G-SchNet}, and FTCP~\cite{FTCP} exhibit substantially larger deviations, with mean RMSD values exceeding 
1.2~\AA~\cite{mattergen}. 
More recently, the transformer-based CrystalFormer~\cite{CrystalFormer} model reported a mean RMSD of 0.96 Å, while random 
structure generation using PyXtal~\cite{pyxtal} yielded an even higher mean RMSD of 1.74 Å. Within this broader landscape, 
DiffCrysGen is positioned between 
the most precise models and weaker generative baselines—thereby demonstrating a favorable balance between structural
 fidelity and model simplicity. This establishes the promise of our diffusion-based approach for realistic structure generation, while 
 highlighting opportunities for further improvement in fidelity.


Most crucially, DiffCrysGen exhibits a substantially higher generation rate, producing structures several orders of magnitude faster than competing diffusion 
models. This improvement arises from its reduced architectural complexity and streamlined sampling procedure. Eventually, the performance of a model 
should be measured by the rate at which it generates stable, unique and novel (S.U.N.) structures. And it is clear that DiffCrysGen has EGR 2—3 orders of 
magnitude higher than MatterGen and DiffCSP. Thus, DiffCrysGen demonstrates a clear, quantitative advantage, yielding a significantly higher rate of 
physically valid and thermodynamically stable structures per unit time.
This shifts the evaluation of generative models from asking “which method is most accurate?” to “which method produces the largest number of physically 
meaningful and usable structures per unit time?” Under this throughput-aware metric, DiffCrysGen generates approximately 
907× more  S.U.N. structures per unit time than MatterGen, and approximately 76× more than DiffCSP (Fig.~\ref{benchmark_plot}(h)), representing significant 
improvement in practical design efficiency.


\subsection*{Evaluation of the generated materials}

\begin{figure}
    \centering
    \includegraphics[scale=0.2]{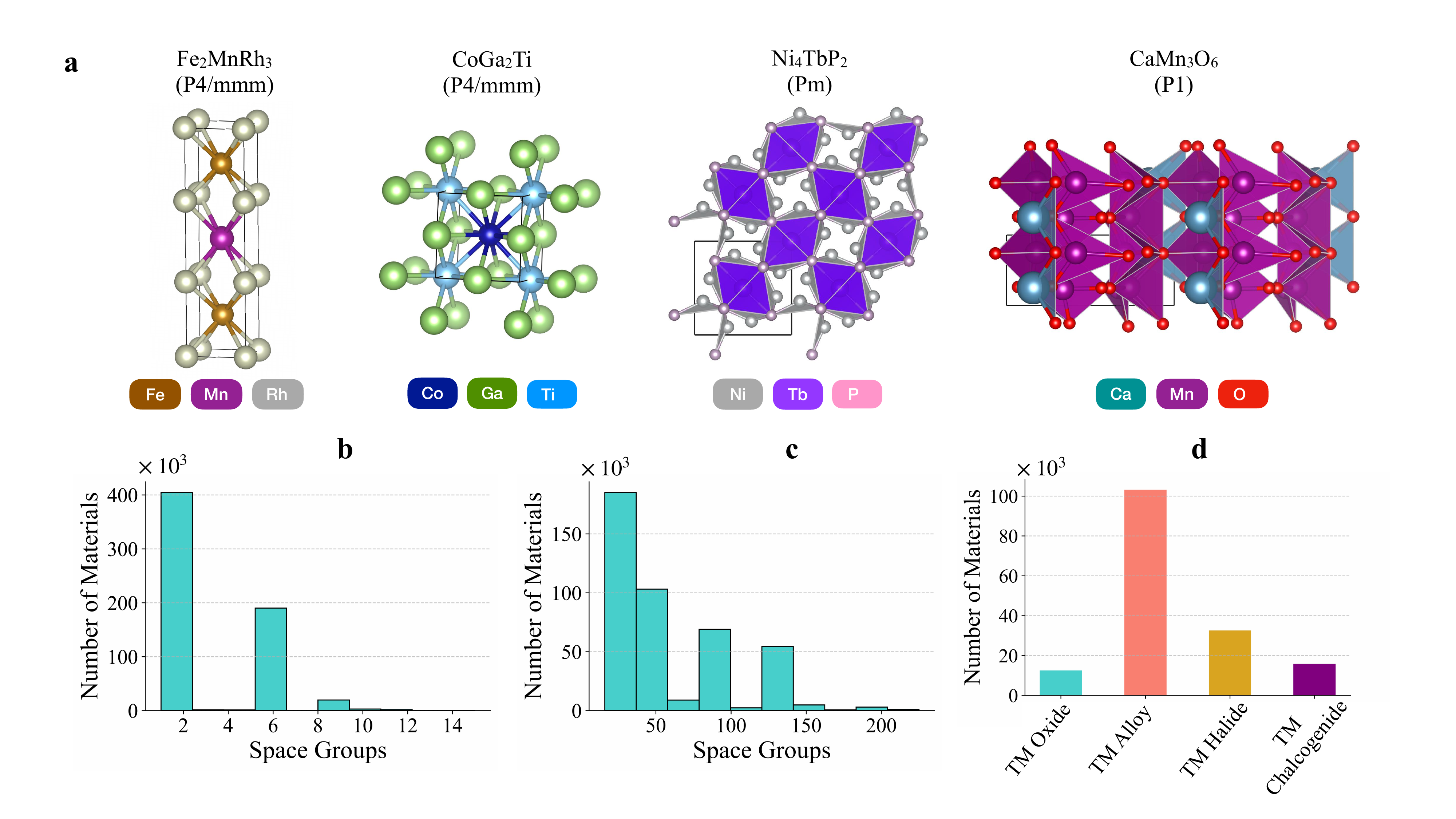}
    \caption{{\bf Generating diverse inorganic crystalline materials}. {\bf a} Visualization of four randomly selected V.U.N. materials generated by DiffCrysGen, with corresponding reduced formula and space group. {\bf b} Distribution of space groups among V.U.N. materials in the triclinic and monoclinic crystal systems. {\bf c} Distribution of space groups among V.U.N. materials in high-symmetry crystal systems (space group number $\ge$ 16). {\bf d} Distribution of different chemical compositions within the RE-free subset of V.U.N. materials containing transition metal (TM) elements.}
    \label{evaluating-generated-materials}
\end{figure}

We generated 1.3 million crystal structures using DiffCrysGen.  spglib (symprec=0.1)~\cite{spglib} could successfully assign space groups to $98.03\%$ of the these. For the remaining $1.97\%$, the assignment of the space group failed, and upon detailed inspection, these failures were due to the presence of overlapping atoms within the unit cell.


$84.80\%$ of the generated structures are valid based on minimum pairwise interatomic distance as defined in the earlier section. We then evaluated the uniqueness and novelty of the valid materials. 


We find that $81.21\%$ of the generated materials (approximately 1.06 million) are V.U.N. (valid, unique, and novel). Among these V.U.N. materials, $16.16\%$ are binary compositions and $83.07\%$ are ternary compositions, showcasing a diverse range of chemical complexities. Fig.~\ref{evaluating-generated-materials}a shows several randomly selected V.U.N. materials, each exhibiting characteristic coordination environments commonly found in inorganic crystals.

Here we want to highlight that we have considered V.U.N. instead of previously defined S.U.N. metric. It is because calculating stability rate for all the 1.3 million structures 
within a reasonable time is computationally prohibitive with the resources we have. 

To assess the structural diversity of the V.U.N. materials, we analyzed their space group distribution. Notably, a substantial $41\%$ exhibit high-symmetry, with space group numbers greater than 16, corresponding to the orthorhombic, tetragonal, trigonal, hexagonal, and cubic systems. Distribution of the low-symmetry and high-symmetry structures are shown in Fig.~\ref{evaluating-generated-materials}b \& c respectively. Existence of $41\%$ high-symmetry structures represents a significant improvement over earlier VAE- or GAN-based generative models, which predominantly produced triclinic structures (over $99\%$), as highlighted in a previous work~\cite{PGCGM}. Only $38\%$ of the V.U.N. materials have the trivial space group P1. Overall, $59\%$ of the V.U.N. materials fall within the low-symmetry triclinic and monoclinic systems, as detailed in Fig.~\ref{evaluating-generated-materials}b. 

In our previous work, the VAE-based model MagGen~\cite{MagGen}, trained on 40,048 materials, exhibited a strong bias toward generating  low-symmetry structures: $99.99\%$ of the generated materials were either monoclinic or triclinic, with $94.49\%$ crystallizing in the trivial P1 space group. To evaluate whether the observed low-symmetry bias stemmed from the limited size of the training dataset, we retrained MagGen using the same, significantly larger dataset employed to train DiffCrysGen. The results remained practically unchanged: $99.82\%$ of the generated materials were still monoclinic or triclinic, and $94.10\%$ belonged to P1. This persistent low-symmetry bias, despite the expanded training data, highlights a fundamental limitation of the VAE architecture, 
and underscores clear advantage of DiffCrysGen's unified diffusion-based approach in capturing and generating materials with diverse crystallographic symmetries.

In the image domain, diffusion models have consistently outperformed VAEs and GANs in generating high-fidelity outputs. Based on the above analyses, we demonstrate that this advantage extends to materials science, particularly in capturing structural symmetries. 

\subsection*{Search for RE-free magnets in DiffCrysGen's generation space}

\begin{figure}
    \centering
    \includegraphics[scale=0.38]{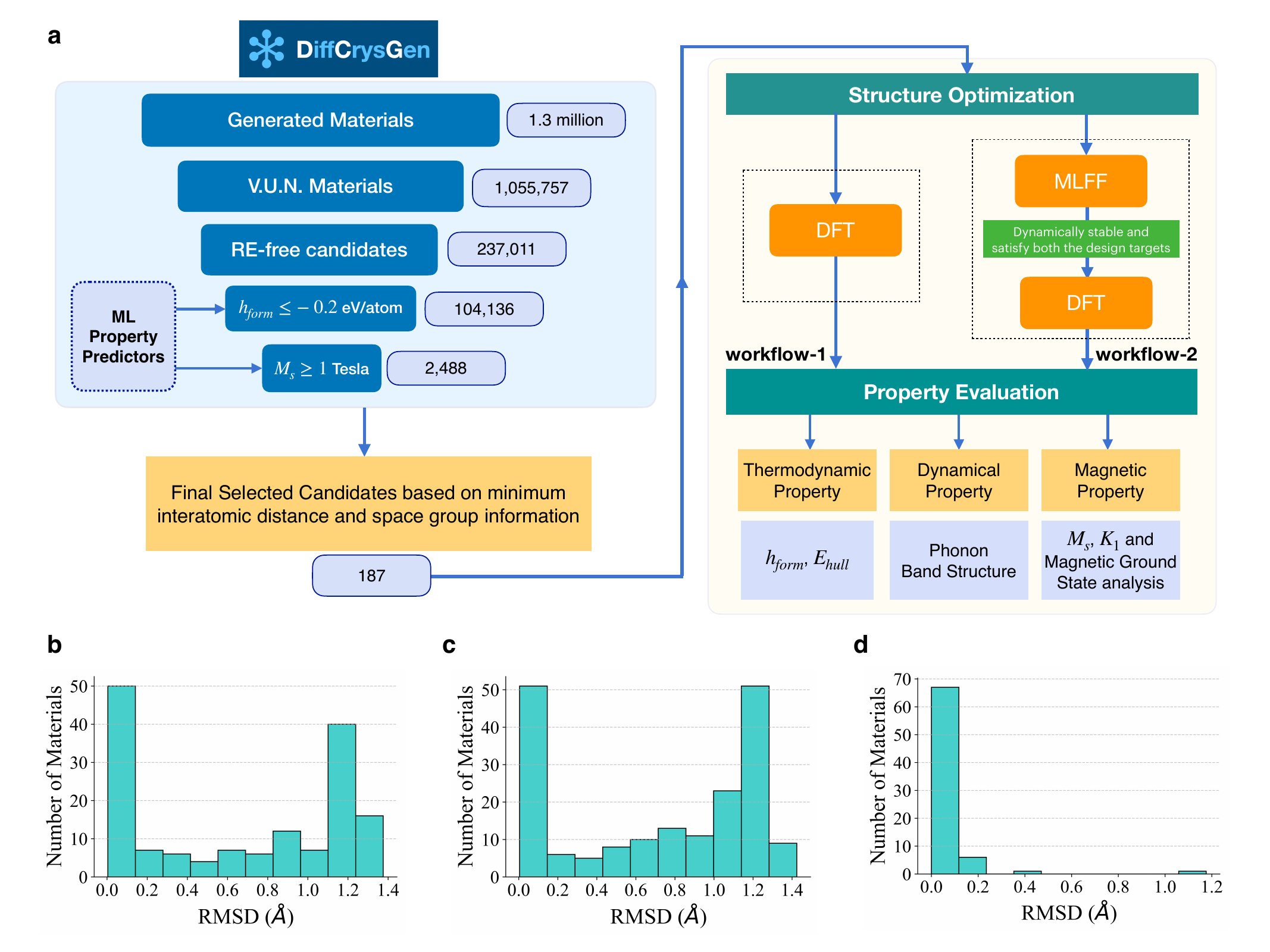}
    \caption{{\bf Materials design pipeline and and structural fidelity assessment}. {\bf a} Hierarchical screening pipeline applied to crystal structures generated by DiffCrysGen for identifying RE-free magnetic materials with high saturation magnetization. The filtering process selects valid, unique, and novel (V.U.N.) compounds without RE elements ($Z_{max}\le54$), followed by screening based on formation energy ($h_{form}$) and saturation magnetization ($M_s$) as predicted by machine-learning surrogate models. The shortlisted candidates are refined through two complementary workflows: DFT (workflow-1) and combined MLFF-DFT (workflow-2) for structural optimization. The final set of materials is then evaluated with DFT to confirm their thermodynamic, dynamical, and magnetic properties.  
    {\bf b} Distribution of root-mean-square deviation (RMSD) between structures generated by DiffCrysGen and those optimized with DFT. {\bf c} Distribution of RMSD between DiffCrysGen-generated and MLFF-optimized structures. {\bf d} Distribution of RMSD between MLFF-optimized and DFT-optimized structures.
   }
    \label{schematic-workflow}
\end{figure}

Since DiffCrysGen generates a wide variety of inorganic materials, we asked if it can help design RE-free magnetic materials with high saturation magnetization ($M_s$). RE-free materials are promising candidates for sustainable permanent magnet applications. For technological applications, such materials are required to have a saturation magnetization $M_s > 1$~Tesla (as is the common practice, we multiply saturation magnetization obtained in units of $\mu_B$ per \AA$^3$~ by the vacuum permeability 
$\mu_0$, and express it in Tesla), and high
magnetic anisotropy energy (MAE) with the anisotropy constant $K_1 > 1$~MJ/m$^3$~\cite{coey}.

To address the question, first we select the subset of the V.U.N. materials that do not contain any RE elements. Specifically, we selected materials having all constituent elements with atomic number $Z_{max}\le54$. This RE-free subset contains a total of 237,011 materials with diverse chemistry, including transition metal (TM) oxides, alloys, halides, and chalcogenides. Distribution of these different chemical compositions is given in Fig.~\ref{evaluating-generated-materials}d.

These RE-free materials were then evaluated using pre-trained machine learning (ML) property prediction models for formation energy ($h_{form}$) and saturation magnetization ($M_s$) (see Methods section for details). We used $h_{form}$ as a key proxy for material stability. Model predictions revealed that $70.67\%$ of these materials are likely stable with negative $h_{form}$, and a more restricted subset of $43.94\%$ have $h_{form}\le-0.2$ eV/atom. Within the latter subset, containing 104,136 materials, we identified 2,488 candidates predicted to possess high saturation magnetization $M_s\ge1$~T. These 2,488 materials represent highly promising candidates for further investigation as potential sustainable permanent magnets. 
Subjecting this entire set to first-principles DFT calculations though not impossible, but is computationally expensive. We implemented additional hierarchical filtering criteria based on our domain knowledge of materials to arrive at a much smaller subset for a more rigorous analysis.

First, to further ensure structural stability and chemical plausibility, we selected materials with a minimum interatomic distance of 1~\AA, which reduced the candidate pool to 1226 materials. We then pruned this dataset by prioritizing ternary materials that belong to high-symmetry crystal classes (space group numbers greater than 16). This specific focus on high-symmetry materials, particularly those within the orthorhombic, tetragonal, and hexagonal crystal systems, was motivated by our previous statistical analysis~\cite{JMMM} of magnet material datasets.
We had seen a strong correlation between these symmetries and high uniaxial magnetic anisotropy ($K_1>0$). This comprehensive filtering process in the present case yielded a final set of 187 highly promising candidates. This hierarchical screening approach, coupled with our computational workflow, is shown schematically in Fig.~\ref{schematic-workflow}a.

\subsection*{Structure Optimization and Property Evaluation}

Having identified this small set of 187 materials generated by DiffCrysGen as the likely candidates for RE-free permanent magnets, we subject
them to two independent workflows, and subsequently merge these to arrive at the most promising materials. Since the structures generated by DiffCrysGen generally not at their equilibrium, the most important step is to relax these to the nearest local minima on the potential energy surface (PES). The first workflow (workflow-1)
involves relaxing the generated structures via DFT, and then calculating their relevant properties. In the second workflow
(workflow-2) we made use of machine learning force fields (MLFFs), which have recently emerged as a powerful alternative to direct DFT relaxations, delivering near-DFT accuracy at a fraction of the cost. While there were no {\it a priori} justifications for employing workflow-2 other than curiosity as workflow-1 already uses DFT, it turns out, {\it a posteriori}, that the two workflows 
complement each other. In some cases they lead to two different polymorphs, one of which would have been missed had we followed only one workflow.

To quantitatively assess the performance of DiffCrysGen and the integrated property prediction pipeline in generating materials that meet preferred design targets, we used the following two key metrics~\cite{FTCP, MagGen} which have been used in the literature. 

(i) DFT Validity rate: The proportion of materials for which structural relaxation converges successfully, relative to the total number of materials subjected to DFT calculations.

(ii) DFT Success rate: The proportion of materials that meet the specified design criteria after DFT relaxation, relative to the total number of materials subjected to DFT calculations.

A successful DFT relaxation confirms both structural and chemical validity, ensuring that the resulting crystal structures exhibit physically meaningful bond lengths, angles, and coordination environments. Importantly, the success rate goes beyond mere structural feasibility. It quantifies how effectively DiffCrysGen can generate novel, stable materials that also meet targeted functional properties.

\subsection*{Workflow-1: DFT}

We optimized the structures of all 187 selected candidates using DFT. Structural relaxation was successful for 155 materials, corresponding to a DFT validity rate of $82.9\%$. To assess their structural fidelity, {\it i.e.}, to quantify how closely the generated structures resemble their equilibrium counterparts, we computed the root-mean-square deviation (RMSD) between the unoptimized and the DFT-optimized atomic positions. The distribution of these RMSD values is shown in Fig.~\ref{schematic-workflow}b. On an average, the generated structures deviate from the DFT-relaxed local minima by 0.65~\AA. We also note that 47 structures ($30\%$ of the optimized set) exhibit RMSD values below 0.1~\AA, indicating that DiffCrysGen often generates configurations already very close to their equilibrium geometries.


Formation energies were calculated after successful structure relaxation. Out of 155, 145 materials exhibit negative formation energies, and 139 meet the target $h_{form}\le-0.2$ eV/atom. A different but overlapping set of 138 materials satisfy the target saturation magnetization ($M_s\ge1$ T). Strikingly, as many as 127 candidates satisfy both design targets simultaneously, giving DFT success rates of $74.3\%$ for $h_{form}$, $73.8\%$ for $M_s$, and an overall DFT success rate of $67.9\%$ for both targets taken together. 

To estimate thermodynamic stability beyond $h_{form}$, we computed distance from the convex hull ($E_{hull}$) for the 127 materials that satisfied both design targets. 
83 of these are found to lie within 0.4 eV/atom of the convex hull, including 30 materials within 0.1 eV/atom. This corresponds to a DFT stability rate $23.6\%$ among these 127 materials, where the stability rate is defined as the percentage of materials lying within 0.1 eV/atom from the hull. 

\subsection*{Workflow-2: MLFF-DFT}

In this workflow, we employed the MatterSim MLFF~\cite{mattersim} to relax the generated structures. Out of the 187 materials, 182 could be successfully optimized. 
The distribution of RMSD between the unoptimized and MLFF-optimized structures is shown in Fig.~\ref{schematic-workflow}c. The mean RMSD is $0.72$~\AA, with 47 materials having RMSD below 0.1~\AA. 

After successful structure optimization with MLFF, magnetic and thermodynamic properties of these 182 materials were calculated using DFT (no structure relaxation).
Out of the 182 MLFF-converged structures, 174 have negative formation energy, 166 exhibit $h_{form}\le-0.2$ eV/atom and 154 materials possess $M_s\ge1$T. A total of 139 materials simultaneously meet both design targets of $h_{form} \le -0.2$~eV/atom and $M_s > 1$~T.

Use of MLFF gives us an inexpensive avenue for screening materials for their dynamical stability via phonon calculations. 
Since phonon calculations in DFT are computationally demanding, we reserve these for later, to be performed on a yet smaller set of materials. 
Out of 139 materials, 78 were found to be dynamically stable without any imaginary phonon frequencies.

Although they have come through a few layers of optimization and screening, these 78 materials are still not at the local minima of the DFT PES. Therefore, these structure were optimized further with DFT. 
This step yielded 75 converged structures. We refer to this set as the MS-DFT structures. Mean RMSD between the MS-DFT and MLFF-converged structures is only 0.051~\AA. Distribution of RMSD for all the materials is shown in Fig.~\ref{schematic-workflow}d. The small mean value shows that the local minima on the MatterSim and DFT PESs are quite close, and a
pre-relaxation by an MLFF before a final DFT relaxation can accelerate the screening process substantially.

DFT relaxation simultaneously produced $h_{form}$ and $M_S$ values. Among the 75 MS-DFT structures, 72 meet the design target 
$h_{form}\le -0.2$~eV/atom, and 74 meet the target $M_S\ge1$~T. Overall, 71 materials satisfy both design targets. 
Convex hull analysis further reveals that 23 out of the 71 materials lie within $0.1$~eV/atom from the hull, and 62 materials are within 
$0.4$~eV/atom from the hull. 
Considering the stable materials within $0.1$~eV/atom of the hull, and removing the overlap with workflow-1 yields 19 unique, magnetic materials.

Having described the two workflows,
we are now in a position to highlight the advantage of using two parallel pathways with an example. The generated material $\text{Co}\text{Mn}\text{O}_2$ (smps-792765) had space group 59. 
Direct DFT relaxation (workflow-1) transformed it to a structure having space group 129, whereas the MLFF–DFT workflow (workflow-2) converged it to space group 31. Remarkably, both the structures emerge as low-energy polymorphs close to the convex hull, with $E_{hull}$ of 33 and 17 meV/atom, respectively. The corresponding RMSD between the generated and optimized structures are $1.06$~{\AA} and $1.09$~{\AA} in the two workflows, indicating that the starting configuration was far from equilibrium. This implies
that the generated structures far from any local minimum may converge to different local minima in the two workflows. This is because it is conceivable that DFT and MLFF can produce slightly different forces. Over long trajectories, as is the case for structures far from equilibrium, the two can end up in two different local minima. Another way of viewing this is that the DFT and MLFF PES are slightly different, particularly far from the minima. 

In contrast, when a generated structure is already close to equilibrium, both workflows converge to nearly the same minimum. For example, $\text{CaMn}_3\text{O}_4$ (smps-1144884), initially in space group 123, relaxed to space group 221 under both workflows, with an RMSD of only $0.01$~{\AA} in both workflows. 

Thus, this dual-path approach not only improves robustness and efficiency of the procedure, but also broadens the accessible landscape of low-energy polymorphs, underscoring the synergistic roles MLFF and DFT can play in accelerating AI-driven materials design.

\subsection*{Final Candidate Pool}

From workflow-1, we selected all materials that satisfy both design targets of $h_{form}$ and $M_s$. To avoid missing promising candidates that could still be realized experimentally, we included metastable phases within $0.4$~eV/atom of the convex hull, a range that remains accessible through non-equilibrium synthesis routes. This yielded a total of 83 materials.

From workflow-2, we restricted the selection to unique structures that are within $0.1$~eV/atom of the convex hull, and that simultaneously meet the design targets, resulting in 19 materials. Although workflow-2 also yields metastable phases (between 0.1 and $0.4$~eV/atom from the hull) that may not appear in workflow-1, we defer their detailed analysis to a future work. Our intention here is to make workflow-2 complementary to workflow-1 for the stable materials. workflow-1 already produces a broad selection of metastable phases (up to $0.4$~eV/atom). workflow-2 is focused on capturing the most promising stable candidates that may have been missed by direct DFT optimization. 

By combining both workflows, we arrive at a final pool of 102 promising candidates, spanning both metastable and thermodynamically stable regimes. All these candidates exhibit large saturation magnetization ($M_s \geq 1$ T). Beyond magnetization, however, a permanent magnet must also display significant constant $K_1$, which stabilizes the magnetization direction and underpins coercivity. We therefore computed $K_1$ for all the 102 materials using DFT. Remarkably but serendipitously, $80.7\%$ exhibit the necessary uniaxial anisotropy, $K_1 > 0$.

Some details need to be discussed at this point. So far, the implicit assumption has been that all the materials are ferromagnetic. But this need not be true as materials may have other magnetic ground states. $K_1$ is evaluated in the FM state before a thorough search for the magnetic ground state is performed, and dynamical stability is established. This strategy is motivated by two considerations. First, computing $K_1$ in the FM configuration provides an efficient screening step that allows us to identify candidates with high uniaxial anisotropy, and to search for their possible AFM states. Second, $K_1$ in the FM configuration serves as a reliable estimate for the intrinsic anisotropy of the material in any magnetic configuration. Since anisotropy is largely dictated by symmetry and spin–orbit interactions, as long as the AFM state does not induce significant structural changes, or change in the atomic moments, $K_1$ is not expected to be much different.

We subjected all thermodynamically stable and metastable materials ($E_{hull} \leq 0.4$ eV/atom) to a dynamical stability check via DFT since their proximity to the convex hull implies enhanced experimental synthesizability, even if their initial anisotropy was modest.
Their properties may be further fine-tuned via doping or strain engineering.
Energies of possible AFM configurations were searched for all materials up to $0.1$~meV/atom from the hull, and those with $K_1 > 1$~MJ/m$^3$
all the way up to $0.4$~meV/atom from the hull.
For materials with AFM configurations lower in energy than FM, we re-examined dynamical stability in the AFM state, and recalculated $K_1$, consistently adopting the dynamically stable lowest-energy state as the true magnetic ground state.

By combining (i) the set of high-anisotropy candidates that are dynamically stable and (ii) the set of thermodynamically and dynamically stable candidates near the convex hull, we ultimately identified 14 ferromagnets and 14 antiferromagnets. These materials exhibit large magnetization (FMs) or sublattice magnetization (AFMs), and some of them have large $K_1$ values. Furthermore, we verified the nonmagnetic (NM) configurations of all these 28 materials and found them to be consistently higher in energy than the corresponding magnetic ground states. 
The FMs are listed in Table~\ref{FM-structures} and the AFMs in Table~\ref{AFM-structures}, with all properties reported in their respective ground states.

We cross-referenced all 28 materials against our training set and three major databases, both experimental and DFT-generated: Materials Project~\cite{MP}, ICSD~\cite{icsd}, and OQMD~\cite{OQMD}. For each match at the composition level, we compared space group numbers to verify structural equivalence (See Supplementary Tables 1 \& 2 for details).

Among the 14 ferromagnets, eight share the same composition with entries in these databases. However, only four—CaMnO$_2$ (smps-1180384), AlCo$_2$Fe (smps-47828), Fe$_2$NiSi (smps-70138), and Co$_2$MnSi (smps-579308)—also match in space group. Detailed inspection reveals that CaMnO$_2$, Fe$_2$NiSi, and Co$_2$MnSi are crystallographically identical to the training set structures (apart from a shift of the unit cell), indicating they are successfully regenerated by the model. It is important to point out that the last three materials converge to the same structure as in the training set after relaxation. As generated, they were found to be in space groups different from those in the training set, and therefore were not discarded while screening for V.U.N. materials.

AlCo$_2$Fe, despite sharing the same lattice and space group, exhibits different site occupations, and thus constitutes a new structure. Also, Fe$_2$NiSi turns out to be a ferromagnet with high uniaxial anisotropy which was previously not reported (we discuss this later in detail). This demonstrates that DiffCrysGen not only generates novel compounds but is also capable of regenerating known materials. 
We have retained all four materials in Table~\ref{FM-structures} since we have studied them in detail, established their dynamical stability and true magnetic ground state.

For the 14 antiferromagnets, eight match at the composition level, but only MgMnO$_2$ (smps-573484) and MnZnO$_2$ (smps-113421) share the same space group with training set entries. Upon detailed inspection, they are found to be crystallographically the same structure. Again, note that they converge to the same structures as in the training set only after relaxation. When generated, they were found to be in different space groups than in the training set.

These were previously studied only in their FM states; we identify their true AFM ground states and establish their dynamical stability, and hence included them in Table~\ref{AFM-structures}.

 In terms of composition, the list contains TM oxides, alloys, and a couple of TM-Si materials. There are both FMs and AFMs among the oxides and alloys. Indeed, identification of metallic AFMs among the generated materials is an interesting development. Metallic AFMs are currently sought after due to their potential application in spintronic devices~\cite{siddiqui}. Mn$_2$Rh$_3$Ti is a particularly interesting metallic AFM due to its large uniaxial anisotropy. LiFe$_2$O$_2$ is another AFM material with large uniaxial anisotropy. Among the FMs, Fe$_2$ZnO$_3$ turns out to have a very large uniaxial anisotropy. LiFeO and ScFe$_4$O$_5$ are also fundamentally interesting materials.

\begin{table}[ht]
    \centering
     \caption{{\bf 14 Ferromagnetic (FM) materials mentioned in text.} All materials are dynamically stable and exhibit high saturation magnetization ($M_s \ge 1$ Tesla). Listed quantities are formation energy ($h_{form}$), convex hull distance ($E_{hull}$), saturation magnetization ($M_s$), and magnetocrystalline anisotropy constant ($K_1$), with units of eV/atom, eV/atom, Tesla, and MJ/m$^3$, respectively. (*) indicates compounds where AFM ordering is lower in energy than FM but dynamically unstable, and thus FM is considered the ground state. Compounds are ordered by $E_{hull}$ in ascending order. Entries with large uniaxial anisotropy ($K_1 \ge 1$ MJ/m$^3$) are highlighted in bold.
     }
     \vspace{0.2cm}
    \renewcommand{\arraystretch}{1.3}  
    \scriptsize
    \begin{tabular}{|c|c|c|c|c|c|c|c|c|}
    \hline
    {\bf ID} & {\bf Composition} & {\bf Crystal Type} & {\bf Space Group} & $\bm{h_{form}}$ & $\bm{M_s}$ & $\bm{E_{hull}}$  & $\bm{K_1}$ \\
    \hline
    smps-711083* & $\text{Mn}_4\text{Ni}\text{O}_5$ & Orthorhombic & Immm (71) & -1.885 & 1.906 & 0 & 0.018 \\
    \bf smps-1262993* & \bf $\text{Fe}_2\text{Zn}\text{O}_3$ & \bf Tetragonal & \bf I4/mmm (139) & \bf -1.628 & \bf 1.47 & \bf 0 & \bf 6.848 \\
    smps-1180384* & $\text{Ca}\text{Mn}\text{O}_2$ & Tetragonal & P4/mmm (123) & -2.666 & 1.143 & 0.002 & -0.029 \\
     smps-142900* & $\text{Co}_3\text{Ni}\text{O}_4$ & Monoclinic & Pm (6) & -1.228 & 1.311 & 0.029 & 0.813 \\
    smps-47828 & $\text{Al}\text{Co}_2\text{Fe}$ & Tetragonal  & P4/mmm (123) & -0.319 & 1.228 & 0.039 & 0 \\
    smps-792162 & $\text{Al}_2\text{Co}_3\text{Fe}_3$ & Orthorhombic & Pmmm (47) & -0.281 & 1.17 & 0.048 & 0.55 \\
    \bf smps-70138 & \bf $\text{Fe}_2\text{Ni}\text{Si}$ & \bf Tetragonal  & \bf P4/mmm (123) & \bf  -0.292 & \bf 1.126 & \bf 0.055 & \bf 1.09 \\
    smps-1099935 & $\text{Fe}_2\text{Rh}\text{Ti}$ & Tetragonal & P4mm (99) & -0.335 & 1.074 & 0.06 & -1.179 \\
    smps-579308 & $\text{Co}_2\text{Mn}\text{Si}$ & Tetragonal  & P4/mmm (123) & -0.351 & 1.299 & 0.073 & 0 \\
    smps-56961 & $\text{Mn}\text{Ni}\text{O}_3$ & Orthorhombic & Imm2 (44) & -1.638 & 1.12 & 0.09 & -0.24 \\
    smps-792765* & $\text{Co}\text{Mn}\text{O}_2$ & Orthorhombic & Pmn2$_1$ (31) & -1.562  & 1.85 & 0.09 & 0.516 \\
    smps-824124 & $\text{Al}\text{Fe}_3\text{Pd}_4$ & Orthorhombic & Pmmm (47) & -0.261 & 1.01 & 0.091 & 0.44 \\ 
    \bf smps-924873 & \bf $\text{Sc}\text{Fe}_4\text{O}_5$ & \bf Orthorhombic & \bf Pmn2$_1$ (31) & \bf -1.941 & \bf 1.65 & \bf 0.198 & \bf 1.91 \\ 
    \bf smps-251736 & \bf $\text{LiFeO}$ & \bf Tetragonal & \bf P4/mmm (123) & \bf -1.272 & \bf 1.11 & \bf 0.376 & \bf 4.20\\

    \hline
    \end{tabular}
    \label{FM-structures}
\end{table}

\begin{table}[ht]
    \centering
    \caption{{\bf 14 Antiferromagnetic (AFM) materials mentioned in the text.} All materials are dynamically stable and exhibit high sublattice magnetization. Listed quantities are formation energy ($h_{form}$), convex hull distance ($E_{hull}$), and magnetocrystalline anisotropy constant ($K_1$), with units of eV/atom, eV/atom, and MJ/m$^3$, respectively. Compounds are ordered by $E_{hull}$ in ascending order. Entries with large uniaxial anisotropy ($K_1 \ge 1$ MJ/m$^3$) are highlighted in bold.
     }
     \vspace{0.2cm}
    \renewcommand{\arraystretch}{1.3}  
    \scriptsize
    \begin{tabular}{|c|c|c|c|c|c|c|c|}
    \hline
    {\bf ID} & {\bf Composition} & {\bf Crystal Type} & {\bf Space Group} & $\bm{h_{form}}$ & $\bm{E_{hull}}$  & $\bm{K_1}$ \\
    \hline
    smps-753026 & $\text{Mg}\text{Mn}_3\text{Pd}_4$ & Orthorhombic & Cmmm (65) & -0.392  & 0 & 0.224 \\ 
    smps-960445 & $\text{Mg}\text{Mn}_3\text{Pd}_4$ & Monoclinic  & P2/m (10) & -0.401  & 0 & -0.433 \\
    smps-573484 & $\text{Mg}\text{Mn}\text{O}_2$ & Tetragonal  & P4/mmm (123) & -2.552  & 0 & 0.019 \\
    smps-439167 & $\text{Mg}\text{Mn}_4\text{Pd}_5$ & Orthorhombic  & Pmmm (47) & -0.383  & 0 & 0.223 \\
    smps-224332 & $\text{Mn}_4\text{Be}\text{Pd}_5$ & Tetragonal & P4/mmm (123) & -0.309  & 0 & 0.10 \\ 
    smps-326567 & $\text{Fe}\text{Ni}\text{O}_2$ & Monoclinic & C2/m (12) & -1.515  & 0.005 & 0.14 \\
    smps-1144884 & $\text{Ca}\text{Mn}_3\text{O}_4$ & Cubic & Pm-3m (221) & -2.344 & 0.006 & 0 \\ 
    smps-914568 & $\text{Mn}_4\text{Na}\text{O}_5$ & Orthorhombic & Immm (71) & -2.027  & 0.016 & 0.38 \\
    smps-792765 & $\text{Co}\text{Mn}\text{O}_2$ & Tetragonal & P4/nmn (129) & -1.599  & 0.033 & 0.22 \\ 
    smps-465027 & $\text{Li}\text{Mn}\text{O}_2$ & Tetragonal  & P4/mmm (123) & -2.110  & 0.048 & -0.61 \\
    smps-113421 & $\text{Mn}\text{Zn}\text{O}_2$ & Tetragonal & P4/mmm (123) & -1.853  & 0.059 & 0.06 \\
    \bf smps-848080 & \bf $\text{Mn}_2\text{Rh}_3\text{Ti}$ & \bf Orthorhombic  & \bf Pmmm (47) & \bf -0.453  & \bf 0.127 & \bf 1.56\\
    \bf smps-93202 & \bf $\text{Li}\text{Fe}_2\text{O}_2$ & \bf Orthorhombic & \bf Immm (71) & \bf -1.384  & \bf 0.252 & \bf 5.12 \\ 
    \bf smps-1246194 & \bf $\text{K}\text{Fe}_2\text{O}_2$ & \bf Orthorhombic & \bf Immm2 (44) & \bf -1.187  & \bf 0.257 & \bf 4.0 \\
    \hline
    \end{tabular}
    \label{AFM-structures}
\end{table}

All our discussions so far on thermodynamic stability of the designed materials have been for 0~K. However, any practical application requires the materials to be
stable at finite temperatures, more specifically, a window of temperature around the room temperature would be of utmost importance. To address this issue,
we have explicitly evaluated the temperature-dependent formation Helmholtz free energy for two of the highlighted ferromagnets 
Fe$_2$ZnO$_3$ and Fe$_2$NiSi. To assess finite-temperature stability, vibrational contributions to the Helmholtz free energy were computed
from the phonon band structure (within the harmonic approximation) post-processed with Phonopy~\cite{phonopy1,phonopy2}. This allows us to 
assess the temperature window of stability of these materials beyond the 0~K convex hull analysis. 
The calculation details are discussed in the methods section. Figure.~S1 in SI shows the temperature-dependent formation Helmholtz free energy $\Delta A(T)$  for 
Fe$_2$ZnO$_3$ and Fe$_2$NiSi. 
Fe$_2$ZnO$_3$ is found to be stable up to 447 K, while Fe$_2$NiSi remains stable up to 473 K. 
Above these temperatures, vibrational entropy favors decomposition into the elemental phases. In this context, it should be clear that 
the formation free energy assesses stability of a compound material with respect to its elemental constituents in their respective reference states, 
and is not a finite temperature
extension of the hull. The latter would require phonon band structure calculations for all the binary and ternary materials used to construct
the hulls in the respective phase spaces. This calls for a larger, community effort.

Another important property of any magnetic material is its transition temperature, and in our context, the Curie temperature ($T_C$) and Néel temperature ($T_N$) for 
FM and AFM materials, respectively. We calculated $T_C$ and $T_N$ for several of the non-oxide FM and AFM materials. The calculations are done within the 
mean-field approximation using the calculated Heisenberg exchange coupling parameters obtained from SPRKKR~\cite{ebert2011calculating}. The computational methodology and technical 
details are discussed in the methods section, and the $T_C$ and $T_N$ are presented in Tables S3 and S4  in the SI. We encountered technical difficulties for the 
materials requiring use of Hubbard $U$ on the TM atoms, and results for these are not reported here. It is interesting to note that the FM material Fe$_2$NiSi and the AFM material 
Mn$_2$Rh$_3$Ti, which have large $K_1$ values, also have transition temperatures 1078~K and 976~K, well above the room temperature. Although mean field methods
are known to overestimate transition temperatures, our experience is that the overestimation is approximately a couple of hundred Kelvins~\cite{jj_prm}.
Given that Fe$_2$NiSi is thermodynamically stable up to $\sim 450$~K, and has a mean-field $T_C$ exceeding 1000~K, 
we can confidently claim that its temperature range of stability is well within the
temperature range over which it remains FM. This makes it attractive for practical applications, and would be an ideal candidate for attempted synthesis.

\subsection*{Understanding some of the interesting materials}

\begin{figure*}
    \centering
    \includegraphics[scale=0.3]{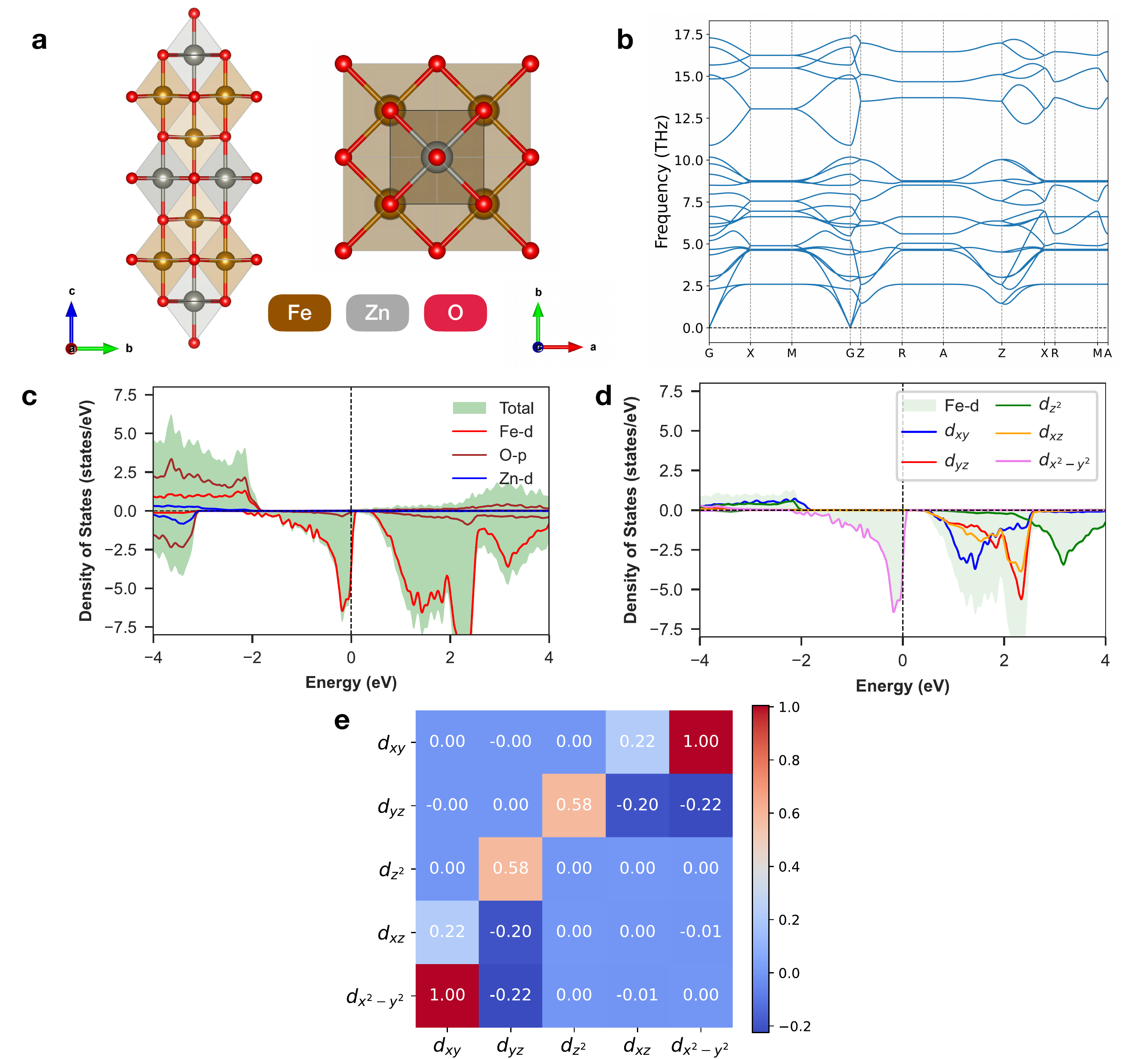}
    \caption{ {\bf Electronic, vibrational, and magnetic properties of $\mathrm{Fe}_2\mathrm{ZnO}_3$.}
{\bf a} Side and top views of the $\mathrm{Fe}_2\mathrm{ZnO}_3$ crystal structure.
{\bf b} Phonon band structure along high-symmetry directions of the Brillouin zone, indicating the dynamical stability of the system. 
{\bf c} Total and atom-projected density of states (DOS). The shaded green region denotes the total DOS, while lines represent projections onto Fe-$d$, O-$p$, and Zn-$d$ orbitals. Positive and negative DOS correspond to the spin-up and spin-down channels, respectively, with the Fermi level ($E_F$) set to 0~eV. 
{\bf d} Orbital-resolved projected DOS (PDOS) for the five Fe-$d$ orbitals ($d_{z^2}$, $d_{xy}$, $d_{yz}$, $d_{xz}$, and $d_{x^2-y^2}$). 
{\bf e} Heatmap of the Fe-$d$ orbital-resolved spin–orbit coupling (SOC) matrix element contributions to the magnetocrystalline anisotropy energy (MAE). The color scale represents the magnitude and sign of the contributions.}
    \label{Fe2ZnO3}
\end{figure*}

\begin{figure*}
    \centering
    \includegraphics[scale=0.3]{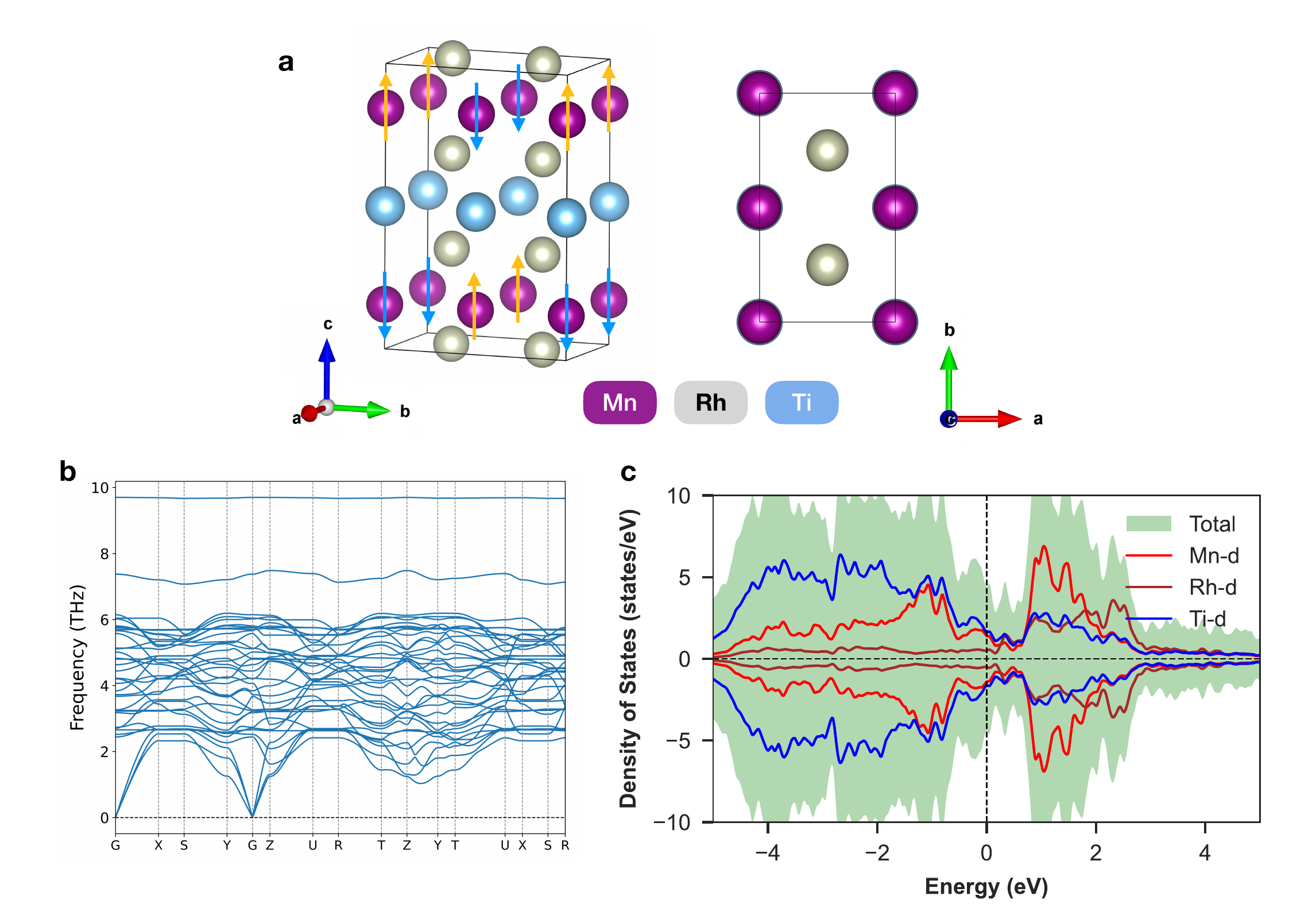}
    \caption{ {\bf Electronic, vibrational and magnetic properties of $\text{Mn}_2\text{Rh}_3\text{Ti}$.}
{\bf a} Side and top views of the $\text{Mn}_2\text{Rh}_3\text{Ti}$ crystal structure.
{\bf b} Phonon band structure along high-symmetry directions of the Brillouin zone, indicating the dynamical stability of the system.
{\bf c} Total and atom-projected density of states (DOS). The shaded green region denotes the total DOS, while lines represent projections onto Mn-$3d$, Rh-$4d$, and Ti-$3d$ orbitals. Positive and negative DOS correspond to the spin-up and spin-down channels, respectively, with the Fermi level ($E_F$) set to 0~eV.
}
    \label{Mn2Rh3Ti}
\end{figure*}

In this section, we analyze electronic and magnetic properties of some of the interesting candidates found in our pipeline. MAE, as calculated above, arises from the spin-orbit coupling (SOC) effects, which is higher for heavy elements, being proportional to $Z^4$, $Z$ being the atomic number of the constituent elements. As pointed out earlier, many of the materials generated by DiffCrysGen, though containing only $3d$ and/or $4d$ elements, have quite high anisotropy. For example, $K_1$ of Fe$_2$ZnO$_3$ is as high as 6.8~MJ/m$^3$, much higher than that of the widely used RE containing permanent magnet $\text{Nd}_2\text{Fe}_{14}\text{B}$ (4.5 MJ/m$^3$). 
Hence, it is interesting and imperative to understand the origin of such large MAE in materials without any heavy elements.

In $3d$ and $4d$ transition-metal systems, SOC energies are typically of the order of a few meV, much smaller than the characteristic electronic (band) energies ($O$(eV)). Therefore, SOC can treated as a perturbation to the non-relativistic (NR) Hamiltonian. 
The SOC Hamiltonian is of the form $H_{LS} = \lambda \mathbf{L} \cdot \mathbf{S}$, where $\lambda$, the SOC constant $\sim Z^4$. The first-order energy correction vanishes by symmetry, and the first non-zero contributions are obtained at the second-order of perturbation~\cite{PhysRevB.47.14932}.

In tetragonal and hexagonal crystal systems, where in-plane rotational symmetry ensures $E_x=E_y$, MAE simplifies to the energy difference between in-plane and out-of-plane magnetization directions, i.e., $\text{MAE} = E_x-E_z$. In this case, to second-order in SOC, 
\begin{equation}
\begin{split}
    \text{MAE}  \approx \lambda^2 
    \sum_{o,u,\sigma} \frac{|\langle o,\sigma | L_z | u,\sigma \rangle|^2 -|\langle o,\sigma | L_x | u,\sigma \rangle|^2}{\epsilon_{u,\sigma} - \epsilon_{o,\sigma}} \\
    +
    \lambda^2 
    \sum_{o,u,\sigma\neq\sigma^{\prime}} \frac{|\langle o,\sigma | L_x | u,\sigma^{\prime} \rangle|^2 -|\langle o,\sigma | L_z | u,\sigma^{\prime} \rangle|^2}{\epsilon_{u,\sigma^{\prime}} - \epsilon_{o,\sigma}}
\end{split}
\label{eq:MAE-perturb}
\end{equation}

For lower-symmetry systems (e.g., orthorhombic), where $E_x\neq E_y$, a full anisotropy characterization would require considering energies along all three crystallographic directions. 

\noindent Here, $| u,\sigma \rangle$ and $| o,\sigma \rangle$ denote 
unoccupied and occupied electronic states with spin $\sigma$, and $\epsilon_{u,\sigma}$, $\epsilon_{o,\sigma}$ are their corresponding energies.
$L_x$ and $L_z$ represent the angular momentum operators along crystallographic $a$ and $c$ axes respectively. 
$Z^4$ dependence of $\lambda$ makes SOC stronger for heavier elements leading to larg anisotropy in RE magnets. However, Eqn.~\ref{eq:MAE-perturb} shows that MAE also depends on the matrix elements of the angular momentum operator connecting the occupied and unoccupied states in same or different spin channels near the Fermi level for materials in which such a perturbative approach is valid. Therefore, even in systems composed of lighter elements, favorable orbital hybridization and electronic structure can lead to large values of MAE. Further consideration shows that, expressed in terms of the atomic $d$-orbitals, only the following SOC matrix elements are non-zero: $\langle d_{yz} | L_x | d_{z^2} \rangle$,
$\langle d_{xy} | L_x | d_{xz} \rangle$,
$\langle d_{yz} | L_x | d_{x^2 - y^2} \rangle$,
$\langle d_{xy} | L_z | d_{x^2 - y^2} \rangle$, and
$\langle d_{yz} | L_z | d_{xz} \rangle$, and the remaining ones vanish identically.

Armed with this information, we arrive at a semi-quantitative  explanation for large magnetic anisotropy in the FM material Fe$_2$ZnO$_3$ using DFT-calculated density of states (DOS) and partial DOS. In addition, we establish metallic AFM character of $\text{Mn}_2\text{Rh}_3\text{Ti}$ by calculating its DOS. Finally, highlight $\text{Fe}_2\text{Ni}\text{Si}$ as another promising candidate for RE-free PMs.

(a) $\text{Fe}_2\text{Zn}\text{O}_3$ has a tetragonal crystal structure (Fig.~\ref{Fe2ZnO3}a) with space group I4/mmm (No. 139). The unit cell consists of a total $12$ atoms. It lies directly on the convex hull, confirming its stability with respect to competing phases. Phonon calculations reveal absence of any imaginary modes (Fig.~\ref{Fe2ZnO3}b), confirming that the structure is dynamically stable. We investigated both FM and AFM configurations. Interestingly, AFM arrangement has a slightly lower energy; but turns out to be dynamically unstable. Therefore, the FM phase is identified as the true ground state. 
The spin-resolved density of states (DOS) (Fig.~\ref{Fe2ZnO3}c) indicate that the system is an insulator.

The magnetic moment in the material arises almost entirely from the Fe sublattice, with each Fe atom carrying a local moment of 3.83 $\mu_\mathrm{B}$. Contributions from Zn and O are negligible. The resulting saturation magnetization ($M_s$) is $1.472$~T.

To elucidate the microscopic origin of the unusually high $K_1$ value, we first examined the atom-resolved spin–orbit coupling (SOC) contributions. The Fe atoms dominate, each contributing $3.364$~meV. This leads to the the uniaxial anisotropy of the material. Further orbital-resolved analysis of the SOC energy differences (Fig.~\ref{Fe2ZnO3}e) reveals that the largest contribution arises from the $\langle d_{xy} | L_z | d_{x^2-y^2} \rangle$ matrix element, $1.51$~meV per Fe atom. A closer look at the orbital-resolved DOS of the Fe atoms in Fig.~\ref{Fe2ZnO3}d  shows that the dominant coupling between these two orbitals has to arise from the same (down) spin channel due to the small gap. First term in Eqn.~\ref{eq:MAE-perturb} then shows that this contribution will be positive, fully justifying our results. 

We have also analyzed the site-resolved orbital angular moments (Table S5 in SI) to provide a microscopic understanding of the large spin-orbit coupling contribution. The results show that Zn and O atoms possess negligible orbital moments for both magnetization directions, whereas the Fe atom carries a significant orbital moment. Notably, the orbital moment of Fe is larger along the crystallographic c-direction than along the a-direction, consistent with the c-axis being the easy magnetization axis. This behavior aligns with Bruno’s model~\cite{Bruno}, which relates magnetocrystalline anisotropy to the anisotropy of orbital moments, predicting that the easy axis corresponds to the direction of larger orbital moment. These results confirm that the large  originates primarily from Fe 3d states and is physically consistent with established SOC-driven anisotropy mechanisms.

A very interesting finding is a large $K_1$, 6.8~MJ/m$^3$, for the 3d transition metal oxide $\text{Fe}_2\text{Zn}\text{O}_3$. 
As is demanded by physics, to take into account electron correlation effects correctly, one needs to use a Hubbard $U$ on the TM $3d$ orbitals.
We used $U= 5.3$~eV on the Fe-$3d$ orbitals, as used in the Alexandria database. However, the exact value of $K_1$ may be sensitive to value of $U$,
which make our conclusions dependent on the particular choice of $U$. To remove this uncertainty, we evaluated the dependence of $K_1$ on $U$ in the range 3-6 eV
a reasonable range of value for Fe $3d$ states. 
$K_1$ decreases to 4.72 MJ/m$^3$ at $U=3$~eV but remains in the range 6.55–6.89 MJ/m$^3$ for $U = 4-6$~eV. 
Thus, while the absolute magnitude of $K_1$ exhibits moderate sensitivity at lower values of  $U$, the predicted large anisotropy remains robust across a 
physically reasonable range of correlation strengths.

Taken together, these results establish $\text{Fe}_2\text{Zn}\text{O}_3$ as a stable ferromagnet with significant uniaxial anisotropy. 
Such a combination of properties makes it a attractive candidate for RE-free permanent magnet applications.

(b) $\text{Mn}_2\text{Rh}_3\text{Ti}$ is a metallic antiferromagnet crystallizing in an orthorhombic structure with space group Pmmm (No. 47), 
as shown in Fig.~\ref{Mn2Rh3Ti}a. Thermodynamically, it lies only $0.127$~eV/atom above the convex hull, placing it within realistic reach of 
experimental synthesis. The structure is dynamically stable, as confirmed by its phonon dispersion (Fig.~\ref{Mn2Rh3Ti}b). Metallic behavior is confirmed 
by the finite DOS at the Fermi level (Fig.~\ref{Mn2Rh3Ti}c). The magnetic moments originate primarily from the Mn atoms, each carrying an absolute 
moment of 3.256 $\mu\mathrm{B}$. Importantly, this material exhibits a large uniaxial 
anisotropy with $K_1 = 1.56$ MJ/m$^3$ with the crystallographic c-axis as the magnetic easy axis.

(c) $\text{Fe}_2\text{Ni}\text{Si}$
is a FM metal in a tetragonal structure with space group P4/mmm (No. 123; Fig. S1(a)). It is thermodynamically stable, lying 55 meV/atom above the convex hull, and also dynamically stable (Fig. S1(b)). The electronic DOS (Fig. S1(c)) confirms its metallic nature. Each Fe atom carries a moment of 2.05~$\mu_{\mathrm B}$, giving a saturation magnetization of $M_s = 1.126$~T. Importantly, $\text{Fe}_2\text{Ni}\text{Si}$ exhibits large uniaxial anisotropy, with $K_1 = 1.09$~MJ/m$^3$ with the crystallographic $c$-axis as the easy axis, and is
estimated to have a Curie temperature well above its temperature range  of thermodynamic stability, the latter extending up to $\sim 450$~K.
These make $\text{Fe}_2\text{Ni}\text{Si}$ another strong candidate for RE-free permanent magnets.

Intriguingly, this material appears in the training set, that is, in the Alexandria database. But its magnetic anisotropy had not been explored. Nominally, a training set material being generated by the model would not be considered for further processing. However, the structure of $\text{Fe}_2\text{Ni}\text{Si}$ generated by DiffCrysGen was found to have space group Pmmm (No. 47), while the one in the training set belongs to P4/mmm (No. 123). That is why it was included for further processing. But after DFT relaxation, it converged to the same structure as the training set sample.
Its successful identification by our generative-screening pipeline highlights the capability of our approach to identify not only novel but also previously overlooked promising candidates.

A polymorph of $\text{Fe}_2\text{Ni}\text{Si}$ in the inverse Heusler structure (Hg$_2$CuTi prototype) and space group F$\bar{4}$3m(No. 216) has been experimentally synthesized~\cite{Luo_2007}. As expected for cubic systems, this phase exhibits negligible magnetocrystalline anisotropy ($K_1 \approx 0$) and lies 41 meV/atom above the convex hull. In contrast, the tetragonal polymorph identified here combines large saturation magnetization and significant uniaxial anisotropy, making it a particularly compelling candidate for experimental realization and further study.


\section*{Discussion}

Diffusion models represent a paradigm shift in the generative design of novel materials. Our study demonstrates that, when trained on sufficiently large datasets, a powerful and expressive diffusion model can inherently learn the complex correlations among atom types, atomic coordinates, and lattice parameters within a single, unified latent space. This approach obviates the need for handcrafted inductive biases or physics-based constraints, thereby simplifying the generative framework.

Our model, DiffCrysGen, exemplifies this capability by successfully generating diverse, novel crystalline materials, which exhibit thermodynamic stability and metastability, and dynamic stability. Even though the generation process was random and unconditional, we are able to identify a number of RE-free magnets, both permanent FM candidates with large saturation magnetization, and metallic AFMs, a testament to the diversity of the generated pool. The efficiency of this design pipeline is particularly noteworthy. By leveraging DiffCrysGen's generative power, alongside our hierarchical ML and DFT filtering, we identified 28 high-performance materials from a pool of 187 generated materials.
We emphasize that to keep the model architecture simple, we did not enforce equivariance to atom permutations, rotations, or reflections, unlike previous approaches~\cite{mattergen,DiffCSP,DiffCSP++,chemeleon}. Despite this, the model achieved competitive performance.

Ultimately, DiffCrysGen is a true game-changer for materials design through its ultra-rapid exploration speed of vast and complex materials spaces, dramatically accelerating the search for novel materials with targeted, high-performance properties.

Despite its remarkable advantages, DiffCrysGen can be improved further. For example, in a small number of cases ($1.97\%$), the model generates materials with overlapping atoms in the unit cell. We believe that incorporating specifically crafted loss functions can more efficiently handle local geometry and resolve this issue.

A recurring challenge for generative models is their inherent bias towards low-symmetry crystal structures. DiffCrysGen has already shown significant progress in overcoming this by generating a large percentage of high-symmetry structures. Further enhancements may be possible through more sophisticated material representations that explicitly encode crystallographic symmetries~\cite{DiffCSP++,WyckoffDiff,SymmCD,WyckoffTransformer}, architectural refinements of the model, and expansion of training datasets. These are the focus of our ongoing research.

In this work, we focused on unconditional material generation to demonstrate the capability of the model to generate diverse inorganic materials. However, the foundational model can be fine-tuned 
to steer the generation process toward specific properties or composition. Such advancements will pave the way for more targeted inverse design in materials science, a direction we are pursuing currently.

\section*{Methods}
\subsection*{Representation of Crystalline Materials}

Crystalline materials are uniquely defined by their unit cell, characterized by its lattice parameters $(a,b,c,\alpha,\beta,\gamma)$, atomic composition, and atomic positions. To systematically encode these components, we employ the Invertible Real-Space Crystallographic Representation (IRCR), a 2D point cloud representation introduced in our previous work \cite{MagGen}.

Briefly, IRCR encodes the unit cell information with five vertically concatenated matrices:
\begin{equation}
\text{IRCR} = \{E,L,C,O,P\},
\end{equation}
where $E,L,C,O,$ and $P$ contain information regarding constituent elements, lattice structure, fractional coordinates, site occupancy, and elemental chemical properties, respectively, summarized in Supplementary Table. S3.  While the property matrix $(P)$ is not directly used for unconditional material generation in this study, it is crucial for property-conditioned design and training predictive models. This representation ensures complete and invertible reconstruction of a material’s lattice, atomic positions, and composition from its encoded data, and offers the flexibility to accommodate diverse compositions and structures. For this work, IRCR was used to encode up to ternary materials.


\subsection*{Diffusion Model }

The essence of generative models lies in modeling the underlying probability distribution of a dataset. Once this distribution is captured, it becomes possible to generate new samples that share similar characteristics with the original data. Here, we implement a score-based diffusion model to learn the data distribution through a stochastic differential equation (SDE)-based framework~\cite{song}. Let the original data be denoted by $\bm{x_0}$, characterized by the unknown complex distribution $p_{\text{data}}(\bm{x_0})$. This data represent the IRCRs of crystalline materials.

\begin{figure*}
    \centering
    \includegraphics[scale=0.2]{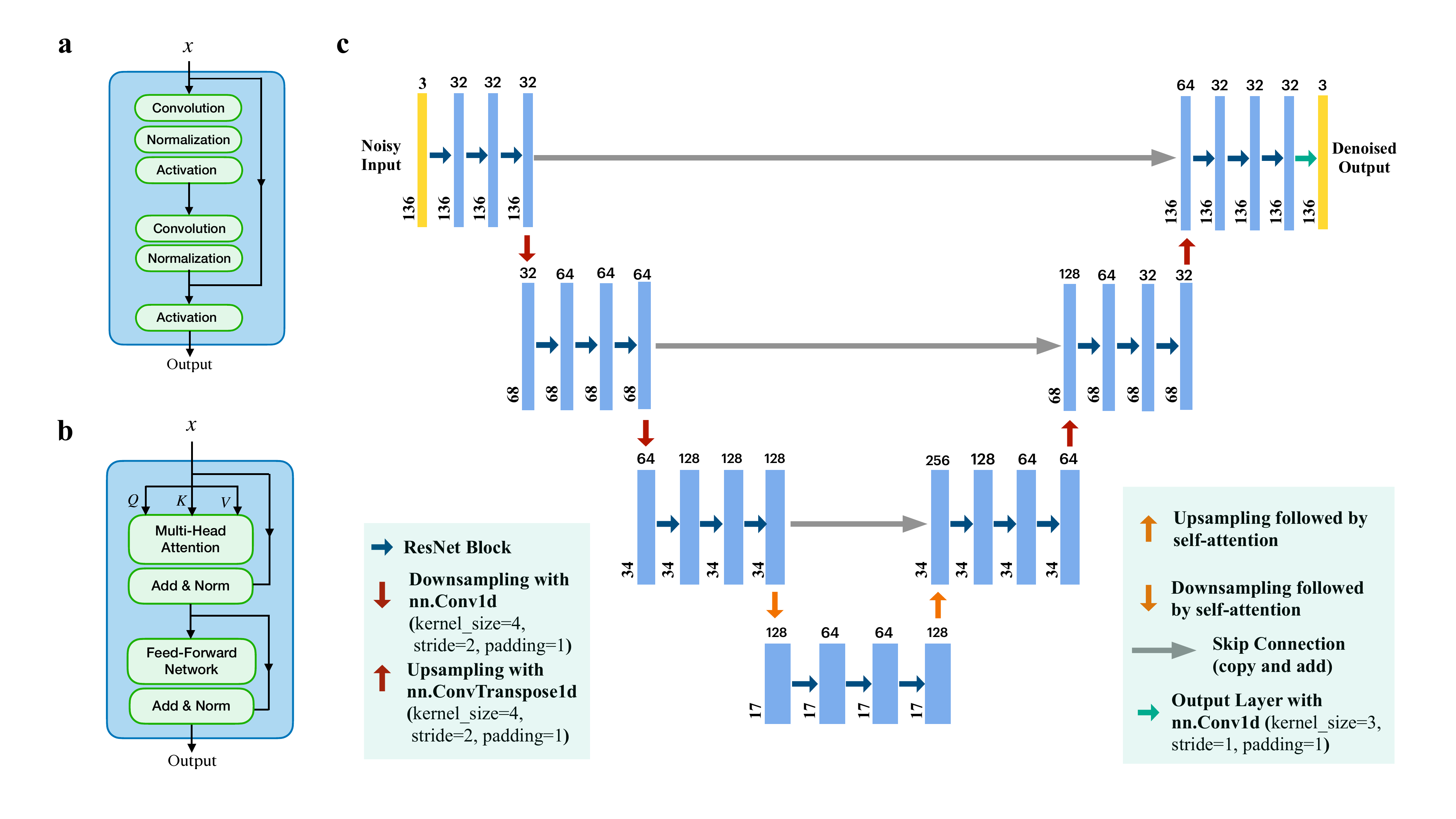}
    \caption{{\bf Architecture of the denoising neural network implemented in DiffCrysGen.} {\bf a} ResNet block, {\bf b} Self-attention block, where $Q$, $K$, and $V$ represent query, key, and value, respectively. {\bf c} UNet architecture, used as the denoising network. Each blue box represents a multi-channel feature map, with the number of channels indicated above the box and the spatial resolution shown at its lower-left corner. Colored arrows denote different operations (e.g., convolution, upsampling, downsampling, skip connection etc.), with color-coding used to distinguish between them.}
    \label{diff-arch}
\end{figure*}

\subsection*{Forward Diffusion Process}

In the forward diffusion process, indexed by a continuous time variable $t \in [0,T]$, we smoothly transform $p_{\text{data}}(\bm{x_0})$ to a known prior distribution $p_T(\bm{x_T})$ by slowly injecting Gaussian noise to the data. This transformation is modeled by the following SDE. 

\begin{equation}
    d\bm{x} =  \sqrt{2\dot{\sigma}(t)\sigma(t)} d\bm{w},
\end{equation}

\noindent where $\sigma(t)$ is the time-dependent noise level and $\bm{w}$ is a standard Weiner process. Here, we implement a variance-exploding (VE) diffusion process by setting $\sigma(t)=t$, leading to a linear increase in variance over time. For this SDE, the perturbation kernel (the conditional distribution of $\bm{x_t}$ given $\bm{x_0}$) can be directly formulated as a Gaussian distribution

\begin{equation}
   p(\bm{x_t}|\bm{x_0}) = \mathcal{N}(\bm{x_t};\bm{x_0},\sigma^2(t)I).
\end{equation}

Consequently, the noisy data $\bm{x_t}$ at any time t is obtained as:
\begin{equation}
  \bm{x_t} = \bm{x_0} + \sigma(t)\bm{\epsilon},
\end{equation}
where $\bm{\epsilon}\sim\mathcal{N}(0,I)$ represents noise sampled from a standard normal distribution.

In this VE setup, the forward diffusion process is governed entirely by stochastic noise. The data diffuses with $\bm{x_0}$ as the mean, while its variance steadily increases with time. By the final step T, the original data structure is fully destroyed and transformed into pure noise with a maximum variance $\sigma_{max}$, yielding the prior distribution $p_T$, which effectively contains no information about the initial data distribution $p_{data}$.

\subsection*{Reverse Diffusion Process}

The reverse diffusion process associated with the forward diffusion describe above is given by an SDE of the form

\begin{equation}
    d\bm{x} = -2\dot{\sigma}(t)\sigma(t)\nabla_{\bm{x}} \log p_t(\bm{x}) dt + \sqrt{2\dot{\sigma}(t)\sigma(t)}d\bm{\bar{w}},
    \label{VE-setup}
\end{equation}

\noindent where $\bar{\bm{w}}$ is the Wiener process defined in reverse time, evolving from $T$ to $0$, and $dt$ represents an infinitesimal negative timestep. Here, $p_t(\bm{x})$ denotes the marginal distribution at time $t$, and $\nabla_{\bm{x}} \log p_t(\bm{x})$ is the corresponding score function. The only unknown in this equation is the score function itself.
By learning an accurate approximation of the score function, we can numerically integrate the reverse SDE Eq.(\ref{VE-setup}) to generate new samples, effectively mapping from the known prior at $t=T$ back to the data distribution at $t=0$.

\subsection*{Probability Flow ODE}

For any given diffusion process described by an SDE, there exists a corresponding deterministic counterpart described by an ordinary differential equation (ODE), known as the probability flow ODE~\cite{song}. The trajectories of this ODE share the same marginal distributions as the original stochastic process at every time step. This property allows the probability flow ODE to serve as a deterministic alternative for sampling from the reverse-time SDE.

For our specific VE diffusion process, the probability flow ODE is given by

\begin{equation}
d\bm{x} = -\dot{\sigma}(t)\sigma(t)\nabla_{\bm{x}} \log p_t(\bm{x}) dt.
\label{ODE}
\end{equation}

Solving this ODE backward in time provides a deterministic path to generate new samples from the data distribution $p_{data}(\bm x_0)$, leveraging the score function.

\subsection*{Score Estimation}

To estimate the score of the marginal distribution, we train a denoiser function $D_\theta(\bm{x_t};\sigma(t))$, implemented as a noise-conditional denoising neural network.
It takes the noisy data $\bm{x_t}$ and the corresponding noise level $\sigma(t)$ as inputs and predicts the clean data $\bm{x_0}$.

The training of the denoiser is guided by minimizing the following loss function
\begin{equation}
    \mathcal{L} = 
    \mathbb{E}_{\bm{x_0}}
    \mathbb{E}_{\bm{x_t}}
    \mathbb{E}_{\sigma}
    [ \lambda(\sigma)
||D_\theta(\bm{x_t};\sigma(t)) - \bm{x_0}||^2_2
    ],
    \label{loss}
\end{equation}
where $\bm{x_0}$, $\bm{x_t}$ and $\sigma$ are sampled from $p_{data}$, $p(\bm{x_t}|\bm{x_0})$ and $p_{train}$, respectively. Here $p_{train}$ specifies the noise level distribution used for training. The weighting function $\lambda(\sigma)$ regulates the contribution of different noise levels during training, ensuring robustness across a wide range of noise levels. After training, the score function $\nabla_{\bm{x}} \log p_t(\bm{x_t})$  can be computed directly from the denoiser using the relation:
\begin{equation} 
\nabla_{\bm{x}_t} \log p_t(\bm{x}_t) =
\frac{D_{\theta}(\bm{x}_t; \sigma(t)) - \bm{x}_t}
{\sigma^2(t)}
\end{equation}

\subsection*{Architecture of the Denoising Network}

The denoiser is implemented as a UNet architecture~\cite{UNet} comprising an encoder and a decoder. The fundamental building block for both components is a ResNet block~\cite{ResNet}, incorporating group normalization~\cite{GroupNorm} and the SiLU activation function throughout. The model utilizes one-dimensional convolutions, with weight standardization applied to each convolutional filter to ensure zero mean and unit variance in filter weights. The detailed structure of the ResNet block is illustrated in Fig.~\ref{diff-arch}a.

At each resolution level within both the encoder and decoder, we employ a sequential stack of three ResNet blocks. Skip connections link corresponding resolution levels between the encoder and decoder. Additionally, a self-attention block~\cite{attention}, implemented using multi-head attention with a single attention head, is incorporated at a feature map resolution of 17. The design of the attention block is depicted in Fig.~\ref{diff-arch}b. The overall UNet architecture is presented in Fig.~\ref{diff-arch}c.

Noise levels are provided to each resolution level in the UNet via sinusoidal positional embeddings, inspired by Transformer models~\cite{attention}, ensuring effective conditioning on the noise scale.

\subsection*{Preconditioning and Training the Denoiser}

To achieve stable and efficient training across the broad dynamic range of noise levels inherent in the diffusion process, we precondition the denoiser. This parameterization follows the framework proposed by Karras et al.~\cite{EDM},

\begin{equation}
D_{\theta}(\bm{x_t}; \sigma) =\ 
 \frac{\sigma^2_{\text{data}}}{\sigma^2 + \sigma^2_{\text{data}}} \bm{x_t} 
 + \frac{\sigma.\sigma_{\text{data}}}{\sqrt{\sigma^2 + \sigma^2_{\text{data}}}} 
F_{\theta}\left(
\frac{\bm{x_t}}{\sqrt{\sigma^2 + \sigma^2_{\text{data}}}},\ 
\frac{1}{4} \ln \sigma
\right).
\end{equation}

Here, $F_{\theta}$ denotes the UNet being trained. The noise level distribution ($p_{train}$) for training is specified as  $ln(\sigma)\sim \mathcal{N}(P_{mean},P^2_{std})$ with $P_{mean}=-1.2$ and $P_{std}=1.2$.
Additionally, $\sigma_{data}$ is set to $0.5$.

The denoising model comprises 1.32 million trainable parameters and was trained for up to 1,000 epochs with a batch size of 64. The training was conducted on a single H100 GPU and four CPU cores, with a requested memory allocation of 16 GB. The total training duration was 16.82 hours.

\subsection*{Generating Samples}

For sample generation, we solve the probability flow ODE in Eq.(\ref{ODE}) using a stochastic sampler~\cite{EDM}, which employs an iterative Heun's 2nd order method to denoise samples from an initial noisy state. Unlike deterministic approaches, it introduces a controlled amount of random noise at each time step, subtly perturbing the sample's trajectory. This inherent stochasticity ensures greater diversity in the final generated samples.

The reverse diffusion timesteps are discretized and geometrically spaced according to the following formula.

\begin{equation}
   t_{i<N} = (\sigma^{\frac{1}{\rho}}_{max} 
   +\frac{i}{N-1}(\sigma^{\frac{1}{\rho}}_{min} - \sigma^{\frac{1}{\rho}}_{max}
   ))^{\rho}, t_N=0,
    \label{time-step}
\end{equation}
where $N$ (the number of reverse diffusion steps) is set to 100, $\rho=7$, $\sigma_{min}=0.002$ and $\sigma_{max}=80$.
We experimented with a larger number of reverse diffusion steps, up to $N=500$, but observed no significant improvement in the quality of generated samples compared to $N=100$. This robust performance with a reduced number of steps $(N=100)$ significantly enhances sampling speed without compromising generative fidelity.

\subsection*{Data Set Construction}

To train DiffCrysGen, we curated a large and diverse dataset from the Alexandria~\cite{Alex-1,Alex-2,Alex-3} database, which contains DFT-computed material properties and crystal structures. The original database comprises 4,489,295 materials. We first filtered structures with $\le20$ atoms per unit cell elemental, binary, or ternary compositions, and elements up to Pu (Z=94), reducing the dataset to 2,584,689 materials.

Next we applied constraints on lattice parameters ($\le25$~\AA), convex hull energy ($E_{hull}\le0.1$~eV/atom), and formation energy ($h_{form}\le0$~eV/atom), yielding 537,049 materials. Since this set included many non-magnetic materials, we further filtered for structures with a saturation magnetization $M_s\ge10^{-5}$ Tesla, resulting in 151,294 materials. We call this curated dataset Alex-1.

The majority of the materials ($94\%$) in Alex-1 are ternary.
The Supplementary Fig. S3(a) shows the distribution of atoms per unit cell in the dataset, revealing that most materials contain more than 10 atoms in the unit cell.
In Supplementary Fig. S3(b) we plot the distribution of space groups, highlighting that the dataset spans all crystal classes. The majority of the materials belong to tetragonal($38.8\%$), orthorhombic($20.2\%$), and monoclinic($15.9\%$) crystal systems. 

Supplementary Fig. S3(c)-(d) illustrate the distributions of $E_{hull}$ and $h_{form}$. $8.8\%$ of the materials lie directly on the convex hull. We have used stable and metastable materials close to the hull to enhance the model's ability to generate energetically viable structures. 

In Supplementary Fig. S3(e) we show the distribution of saturation magnetization ($M_s$). Most materials exhibit relatively low magnetization. Notably, only 21\% of the materials have a saturation magnetization exceeding $0.5$~Tesla. Only a small fraction of materials have significantly high magnetization. This underscores the inherent challenge in designing high-performance magnetic materials.

Finally, we split Alex-1 into an 80:10:10 ratio for training, validation, and testing the diffusion model. Supplementary Fig. S4(a) presents the learning curve, which exhibits stable convergence, with the loss decreasing smoothly and gradually, without abrupt fluctuations, indicating a well-behaved and effective training process.

\subsection*{Property Prediction Models}

We trained two separate property prediction models, built using convolutional neural networks~\cite{CNN}, with IRCR as the input feature: one for $h_{form}$ and another for $M_s$. For $M_s$ prediction, we used the Alex-1 dataset. However, since Alex-1 contains materials with negative $h_{form}$ only, we expanded the dataset by including unstable materials also to improve the robustness of the $h_{form}$ predictor. Specifically, we augmented Alex-1 with 313,670 unstable materials, resulting in a combined dataset of 464,964 materials, which we refer to as Alex-2. This expanded dataset was used exclusively for training the $h_{form}$ regressor. 

The property prediction models exhibit exceptional accuracy, with the formation energy ($h_{form}$) predictor achieving a mean absolute error (MAE) of $0.046$~eV/atom and an $R^2$ score of $0.99$, while the saturation magnetization ($M_s$) predictor attains an MAE of $0.054$~Tesla with an $R^2$ score of $0.92$. We present the parity plots for $h_{form}$ and $M_s$, comparing the actual and predicted values, in Supplementary Fig. S4(b)-(c). The predictions exhibit a well-balanced distribution across the entire property range, demonstrating the robustness of both models. 

While the state-of-the-art graph neural networks trained on the entire Alexandria database~\cite{SCHMIDT2024101560} have achieved lower MAEs ($0.0164$~eV/atom for $h_{form}$ and $0.031$~Tesla for $M_s$), our models demonstrate comparable performance even with a much smaller training data set. This highlights the efficiency of our model, making it a powerful tool for high-throughput screening of novel magnetic materials.

\subsection*{DFT Calculations}

All structure optimizations and total energy calculations were performed using spin-polarized DFT within VASP~\cite{vasp1,vasp2}. The computational parameters were chosen to be consistent with the Alexandria database, which is borrowed from the Materials Project (MP) database. We employed the projector-augmented wave (PAW)~\cite{PAW} method with Perdew-Burke-Ernzerhof (PBE)~\cite{PBE} exchange-correlation pseudopotentials and a plane-wave cutoff energy of 520 eV. Brillouin zone integrations were carried out using uniform  $\Gamma$-centered k-point meshes with a density of $0.15~\text{\AA}^{-1}$. Full structural relaxations were performed, allowing both lattice vectors and atomic positions to vary until the total energy and atomic forces converged below $10^{-5}$ eV and $0.005$ eV/\AA, respectively. 

For transition metal oxides and fluorides, we employed GGA+$U$ calculations following the Dudarev approach~\cite{dudarev}, while standard GGA was used for all other compounds. The Hubbard $U$ parameters were adopted from the Materials Project (MP) database~\cite{MP}, consistent with those used in the Alexandria database (e.g., $U = 5.3$ eV for Fe). To calculate the formation energy, we have included (i) the energy corrections of transition metal atoms for mixing GGA and GGA+U frameworks~\cite{PhysRevB.84.045115}, and (ii) anionic energy correction terms for oxygen and fluorine atoms~\cite{Wang2021}.
Energy from the hull ($E_{hull}$) is calculated using the pymatgen~\cite{pymatgen} PhaseDiagram module with respect to the materials in the MP database.

For the calculation of the magnetocrystalline anisotropy energy constant $K_1$, we performed non-collinear DFT calculations including spin-orbit coupling (SOC) in the Hamiltonian. A denser \textit{k}-point mesh with a sampling density of $0.10$~\AA$^{-1}$, and an energy convergence criterion of $10^{-7}$ eV were used to ensure convergence of the total energy. 

First, we evaluated the total energies with magnetization constrained along the crystallographic a,b, and c axes. 
The lowest-energy orientation defines the easy direction (easy axis or, if a plane of directions is degenerate, easy plane) and the highest energy orientation defines the hard axis.

For systems with three distinct energies ($E_a\neq E_b\neq E_c$), we define the MAE as the smallest reorientation energy barrier relative to the easy direction:
\begin{equation}
\text{MAE} = E_\text{intermediate} - E_\text{easy}, 
\end{equation}
where the direction with the second lowest total energy is the intermediate axis. This choice reflects the first barrier encountered by the magnetization when rotating away from the easy axis. 

For cases where a plane of directions is energetically degenerate but lies at higher energy than the remaining direction (i.e., $E_a=E_b>E_c$), the degenerate plane is taken as the hard direction, and the MAE is defined as
\begin{equation}
\text{MAE} = E_\text{hard} - E_\text{easy}.
\end{equation}

For easy-plane systems, the MAE is taken as 
\begin{equation}
\text{MAE} = -(E_\text{hard} - E_\text{easy}), 
\end{equation}
giving a negative value. 

The first-order anisotropy constant was then extracted as 
\begin{equation}
\frac{\text{MAE}}{V} = K_1 \sin^2\theta,
\end{equation}
where $V$ is the unit-cell volume, and $\theta$ is the angle between the magnetization direction and the easy direction. 
A positive $K_1$ ($K_1 > 0$) indicates easy-axis anisotropy, whereas a negative $K_1$ ($K_1 < 0$) corresponds to easy-plane anisotropy.

Phonon calculations were performed using density functional perturbation theory (DFPT)~\cite{dfpt} with a strict electronic energy convergence threshold of $10^{-8}$~eV. Post-processing was performed using phonopy~\cite{phonopy1,phonopy2}. 

For each material, we determined the magnetic ground state by comparing FM and multiple AFM configurations. The AFM configurations were systematically generated using the MagneticStructureEnumerator module in pymatgen~\cite{pymatgen}, ensuring an exhaustive exploration of symmetry-distinct spin arrangements.


\subsection*{Finite Temperature Stability}

The vibrational Helmholtz free energy is given by
\begin{equation}
A_{\mathrm{vib}}(T)
=
\frac{1}{2}\sum_{q,\nu} \hbar \omega_{q\nu}
+
k_B T \sum_{q,\nu}
\ln \left( 1 - e^{-\hbar \omega_{q\nu}/k_B T} \right)
\end{equation}
where $w_{q\nu}$ denotes the phonon frequency at wave-vector $q$ and branch index $\nu$, and $k_B$ is the Boltzmann constant. The first term corresponds to the zero-point energy (ZPE), while the second term accounts for the temperature-dependent vibrational entropy.
The total Helmholtz free energy of a phase at temperature $T$ was constructed as 
\begin{equation}
A(T) = E_{DFT} + A_{vib}(T),
\end{equation}
where $E_{DFT}$ is the static DFT total energy at 0~K.

The temperature-dependent formation Helmholtz free energy was then evaluated as
\begin{equation}
\Delta A(T) = A_{\text{compound}}(T)-\sum_i x_iA_i(T),
\end{equation}
where the sum over $i$ runs over all the constituent elements of the compound material, $x_i$ are the stoichiometric coefficients and $A_i(T)$ are the Helmholtz free energies of the elemental reference phases. Since vibrational contributions were computed relative to 0 K, the formation free energy can be written as 
\begin{equation}
\Delta A(T) = h_{form}(0\,\mathrm{K}) + \Delta A_{\mathrm{vib,form}}(T),
\end{equation}
with
\begin{equation}
\Delta A_{\mathrm{vib,form}}(T)
=
A_{\mathrm{vib}}^{\mathrm{compound}}(T)
-
\sum_i x_i \,  A_{\mathrm{vib}}^{(i)}(T)
\end{equation}

All quantities were normalized per atom. For crystalline solids at ambient pressure, the $pV$ term is negligible ($10^{-5}$~eV/atom), such that the Helmholtz free energy provides an excellent approximation to the Gibbs free energy.

We also wish to highlight that for oxygen, the reference state was taken as gas-phase O$_2$, consistent with standard thermodynamic conventions for oxide formation reactions. The temperature-dependent thermodynamic properties of O$_2$ were evaluated using the Shomate equation, with fitting parameters obtained from the NIST Chemistry WebBook~\cite{NIST}. This yields the Gibbs free energy for oxygen gas.  The Shomate coefficients employed in this work are valid within the temperature range of 100-700 K.  Accordingly, the temperature-dependent formation Helmholtz free energy (which is essentially equal to the Gibbs free energy, as the $pV$ term is very small) for Fe$_2$ZnO$_3$ was evaluated and plotted only within this temperature window to ensure thermodynamic consistency with the validity range of the NIST Shomate parameters for gaseous O$_2$.

For most practical purposes, one would be interested in a temperature range in the neighborhood of the room temperature. Our analysis shows that both the compounds are stable in this temperature region. The formation free energy plot analyze stability of a compound relative to its constituent elemental phases at finite temperatures. It is not a finite temperature extension of the convex hull, however. The latter would require phonon band structure calculations of all the binary and ternary phases used to construct the hull, and requires serious computational efforts and resources. It can be a part of a worthwhile future community effort to construct more realistic free-energy hulls.


\subsection*{Curie and Néel Temperature Calculations}

The calculation of Heisenberg exchange coupling constant $J_{ij}$ is imperative in exploring the details of the magnetic exchange interactions between the atoms. Spin-polarized relativistic Korringa-Kohn-Rostoker (SPRKKR) code \cite{ebert2011calculating}, has been used to calculate $J_{ij}$ within the real space approach introduced by Liechtenstein and co-workers, referred to as the LKAG formalism \cite{liechtenstein1987local}. All our calculations are done within the scalar relativistic representation of the valence states while, the exchange-correlation potential was modeled using the Perdew-Burke-Ernzerhof (PBE) parametrization \cite{PhysRevLett.77.3865} and the angular momentum expansion up to $l_{max}$ = 3 was taken for each atom. In order to further improve the charge convergence with respect to $l_{max}$,  Lloyd’s formula has been employed for determining the Fermi energy \cite{doi:10.1080/00018737200101268, Zeller_2008}

Considering the classical Heisenberg model, the Hamiltonian of a spin system can be given by 
\begin{equation}
\label{eqn:E5}
H = -\sum_{i \neq j} J_{ij}\, \mathbf{e}_{i}\cdot\mathbf{e}_{j},
\end{equation}
where $J_{ij}$ denotes the Heisenberg pair exchange coupling parameter between sites $i$ and $j$, and $\mathbf{e}_{i}(\mathbf{e}_{j})$ is the unit vector pointing along the direction of the local magnetic moment at site $i$ ($j$).

Within the mean-field approximation (MFA), the Curie temperature $T_{c}^{\mathrm{MFA}}$ for a single-sublattice system is given by
\begin{equation}
\label{eqn:E6}
\frac{3}{2} k_{B} T_{c}^{\mathrm{MFA}} = J_{0} = \sum_{j} J_{0j},
\end{equation}
where $J_{0}$ represents the on-site exchange parameter arising from interactions of a reference atom with all neighboring magnetic atoms.

For a multi-sublattice system, the magnetic moments on different crystallographically inequivalent sites must be treated separately. Hence, one has to solve the following coupled equations
\begin{equation}
\label{eqn:E7}
\frac{3}{2} k_{B} T_{c}^{\mathrm{MFA}} \langle e^{\mu} \rangle
=
\sum_{\nu} J_{0}^{\mu\nu} \langle e^{\nu} \rangle,
\end{equation}
with
\begin{equation}
\label{eqn:E8}
J_{0}^{\mu\nu}
=
\sum_{r \neq 0}
J_{0r}^{\mu\nu}.
\end{equation}

Here, the indices $\mu$ and $\nu$ label the distinct magnetic sublattices, while $r$ denotes the lattice vector specifying the position of atoms within a given sublattice. The quantity $\langle e^{\nu} \rangle$ represents the thermal average of the $z$ component of the unit vector $\mathbf{e}_{r}^{\nu}$ describing the direction of the local magnetic moment at site $(\nu,r)$, and therefore corresponds to the reduced (sublattice-resolved) magnetization. The matrix element $J_{0}^{\mu\nu}$ denotes the total exchange field acting on a reference atom in sublattice $\mu$ due to all magnetic atoms belonging to sublattice $\nu$. The diagonal elements ($\mu=\nu$) account for intra-sublattice exchange interactions, while the off-diagonal elements ($\mu\neq\nu$) describe the inter-sublattice exchange coupling between different magnetic networks.

The coupled mean-field equations can be written in matrix form as
\begin{equation}
\label{eqn:E9}
(\boldsymbol{\Theta} - T \mathbf{I}) \mathbf{E} = 0,
\end{equation}
where $\mathbf{I}$ is the identity matrix and $\mathbf{E}^{\nu} = \langle e^{\nu} \rangle$ is the vector of sublattice-resolved order parameters. The elements of the matrix $\boldsymbol{\Theta}$ are defined as
\begin{equation}
\label{eqn:E10}
\frac{3}{2} k_{B} \Theta_{\mu\nu} = J_{0}^{\mu\nu}.
\end{equation}

The condition for a non-trivial solution of Eq.~\ref{eqn:E9} requires that the determinant of $(\boldsymbol{\Theta} - T\mathbf{I})$ vanishes, which leads to an eigenvalue problem. The solution of the eigenvalue problem can lead us to $T_c$, which is the largest eigenvalue of the $\Theta$ matrix\cite {ANDERSON196399}. The $r$-summation in Eqn.~\ref{eqn:E8} has been carried out for a radius of $r_{max} = 7a$, where $a$ denotes the lattice constant.

\subsection*{Machine Learning Force Field}

MatterSim v1.0.0-5M was employed as the MLFF in workflow-2. Structural optimizations were carried out using the FIRE optimizer in conjunction with the Frechet Cell Filter, as implemented in the Atomic Simulation Environment (ASE)~\cite{ASE}.

At the MLFF level, phonon band structures were computed using the finite-displacement method with $(5 \times 5 \times 5)$ supercells, as implemented in ASE. This setup ensures adequate sampling of the Brillouin zone and provides reliable predictions of dynamical stability prior to DFT refinement.

\subsection*{Software}
All codes were written in Python (v3.11.11). For dataset preparation, we used NumPy (v2.0.1) and Pandas (v2.2.3). The neural network models were implemented in PyTorch~\cite{PyTorch} (v2.5.1). Data normalization was done using scikit-learn~\cite{scikit-learn} (v1.6.1). All the crystallographic information files (CIFs) were processed by atomic simulation environment (ASE-v3.24.0)~\cite{ASE}. Space groups were determined using the spglib library~\cite{spglib} (v2.5.0).

\subsection*{Author contributions}
S. Mal developed the DiffCrysGen model, the property predictions models, performed the DFT-based validation of generated materials (workflow-1), and analyzed the electronic structure and origin of large magnetic anisotropy in selected materials. N. Ahmed performed the combined MLFF and DFT based validation (workflow-2). S. Mal and N. Ahmed performed the comparative benchmarking against other state-of-the-art crystal diffusion models, and prepared the manuscript. J. Jami calculated $T_c$'s and $T_N$'s. S. Mishra suggested and supervised the implementation of the diffusion-based generative model. P. Sen conceived of the project, performed an overall supervision of the project, and also contributed to manuscript writing. All authors have read and approved the manuscript.

\subsection*{Competing interests}
The authors declare to have no competing interests.

\subsection*{Data Availability}
The dataset used to train the generative model and the property prediction models, and the DFT-optimized crystal structures of all final candidate materials highlighted in the manuscript will be made 
available at the GitHub repository: \url{https://github.com/SouravMal/DiffCrysGen}.


\subsection*{Code Availability}
The source code for the generative model developed in this work is available at the GitHub repository: \url{https://github.com/SouravMal/DiffCrysGen}.

\section*{Acknowledgements} 
The work was funded by the DAE, Govt. of India, through institutional funding to HRI. All calculations were performed in the cluster computing facility at HRI~\cite{hri-hpc}.

\bibliographystyle{naturemag}
\bibliography{updated_ref_arxiv}

\end{document}


\title{\Large \textbf{Supplementary Information}}
\subtitle{DiffCrysGen: A Generative Diffusion Model for Accelerated Design of Inorganic Crystalline Materials}

\author[1,2]{\fnm{Sourav} \sur{Mal}}

\author[3]{\fnm{Nehad} \sur{Ahmed}}

\author[1,2]{\fnm{Junaid} \sur{Jami}}

\author[4,2]{\fnm{Subhankar} \sur{Mishra}}\email{smishra@niser.ac.in}

\author[1,2]{\fnm{Prasenjit} \sur{Sen}}\email{prasen@hri.res.in}

\affil[1]{\orgname{Harish-Chandra Research Institute},
\orgaddress{\street{Chhatnag Road, Jhunsi},
\city{Prayagraj},
\postcode{211019},
\country{India}}}

\affil[2]{\orgname{Homi Bhabha National Institute},
\orgaddress{\street{Training School Complex, Anushakti Nagar},
\city{Mumbai},
\postcode{400094},
\country{India}}}

\affil[3]{\orgdiv{Department of Chemistry},
\orgname{Indian Institute of Science Education and Research (IISER) Tirupati},
\orgaddress{\city{Tirupati},
\state{Andhra Pradesh},
\postcode{517619},
\country{India}}}

\affil[4]{\orgdiv{School of Computer Sciences},
\orgname{National Institute of Science Education and Research (NISER)},
\orgaddress{\city{Jatni},
\state{Odisha},
\postcode{752050},
\country{India}}}

\maketitle

\clearpage
\section*{Benchmarking Generated Structures Against Existing Databases}

\begin{table}[htbp]
  \centering
  \caption{{\bf Comparison of space groups for ferromagnetic structures with matching compositions.} We compare the space groups of final DFT‑optimized ferromagnetic structures from DiffCrysGen with those having same composition in our training set and external databases (Materials Project (MP), ICSD, and OQMD). Where the space group from DiffCrysGen (DFT optimized) matches a space group in another source, it is highlighted in red. 
 }
  \label{tab:FM-materials-comparison}
\renewcommand{\arraystretch}{1.2}
\setlength{\tabcolsep}{3.2pt}
\rowcolors{2}{gray!10}{white}
\begin{tabularx}{\textwidth}{|l|l|C|C|L|L|L|}
\hline
\rowcolor{gray!20} \textbf{ID} & \textbf{Composition} & \textbf{DiffCrys Gen} & \textbf{Training Set} & \textbf{MP} & \textbf{ICSD} & \textbf{OQMD} \\
\hline
smps-1180384 & CaMnO\textsubscript{2} & \textcolor{red}{123} & \textcolor{red}{123} & 2, 11, \textcolor{red}{123}, 166, 227, 59, 62 & -- & 11, 31, 227, 26, \textcolor{red}{123}, 62, 63, 166 \\
smps-142900 & Co\textsubscript{3}NiO\textsubscript{4} & 6 & 65 & -- & -- & -- \\
smps-47828 & AlCo\textsubscript{2}Fe & \textcolor{red}{123} & \textcolor{red}{123} & 225 & 225, 221 & 225, 58, \textcolor{red}{123}, 65, 216, 194 \\
smps-70138 & Fe\textsubscript{2}NiSi & \textcolor{red}{123} & \textcolor{red}{123} & 216 & -- & 216, 79, 225 \\
smps-1099935 & Fe\textsubscript{2}RhTi & 99 & 123 & -- & -- & 216, 225 \\
smps-579308 & Co\textsubscript{2}MnSi & \textcolor{red}{123} & \textcolor{red}{123} & \textcolor{red}{123}, 225 & 225, 216 & 225, 58, 139, \textcolor{red}{123}, 74, 216, 194 \\
smps-56961 & MnNiO\textsubscript{3} & 44 & -- & 2, 148 & 148, 36 & 148, 62, 36, 167, 99, 221, 123 \\
smps-792765 & CoMnO\textsubscript{2} & 31 & -- & -- & -- & 12, 225, 166, 216 \\
\hline
\end{tabularx}  
\end{table}

\begin{table}[htbp]
  \centering
  \caption{{\bf Comparison of space groups for antiferromagnetic structures with matching compositions.} We compare the space groups of final DFT‑optimized antiferromagnetic structures from DiffCrysGen with those having same composition in our training set and external databases (Materials Project (MP), ICSD, and OQMD). Where the space group from DiffCrysGen (DFT optimized) matches a space group in another source, it is highlighted in red. 
 }
  \label{tab:AFM-materials-comparison}
  \renewcommand{\arraystretch}{1.2}
\setlength{\tabcolsep}{3.2pt}
\rowcolors{2}{gray!10}{white}
\begin{tabularx}{\textwidth}{|l|l|C|C|L|L|L|}
\hline
\rowcolor{gray!20} \textbf{ID} & \textbf{Composition} & \textbf{DiffCrys Gen} & \textbf{Training Set} & \textbf{MP} & \textbf{ICSD} & \textbf{OQMD} \\
\hline
smps-573484 & MgMnO\textsubscript{2} & \textcolor{red}{123} & \textcolor{red}{123} & 11, 160, 166, 2, 227, 58, 62 & -- & 11, 12, 26, 31, 58, 62, \textcolor{red}{123}, 227 \\
smps-439167 & MgMn\textsubscript{4}Pd\textsubscript{5} & 47 & 123 & -- & -- & -- \\
smps-326567 & FeNiO\textsubscript{2} & 12 & -- & -- & -- & 14, 1 \\
smps-1144884 & CaMn\textsubscript{3}O\textsubscript{4} & 221 & 65 & 123, 65 & -- & 13, 6, 223 \\
smps-792765 & CoMnO\textsubscript{2} & 129 & -- & -- & -- & 12, 225, 166, 216 \\
smps-465027 & LiMnO\textsubscript{2} & 123 & 12, 2, 141, 8, 58, 15, 62, 59, 166 & 1, 12, 141, 15, 156, 166, 18, 2, 25, 58, 59, 62, 74, 8, 9 & 25, 58, 141 & 141, 59, 62, 58, 12, 166, 92, 33, 31 \\
smps-113421 & MnZnO\textsubscript{2} & \textcolor{red}{123} & \textcolor{red}{123}, 156 & 1, 2, 12 & -- & 122, 12 \\
smps-848080 & Mn\textsubscript{2}Rh\textsubscript{3}Ti & 47 & 123 & -- & -- & 164 \\
\hline
\end{tabularx}
\end{table}

\clearpage
\section*{Curie and Néel Temperature of the Selected Materials}
\begin{table}[htbp]
  \centering
  \caption{Curie temperature of the ferromagnetic materials}
  \label{tab:FM-TC}

\renewcommand{\arraystretch}{1.2}
\setlength{\tabcolsep}{3.2pt}
\rowcolors{2}{gray!10}{white}

\begin{tabularx}{\textwidth}{|L|L|L|L|C|}\hline
\rowcolor{gray!20}
\textbf{ID} & \textbf{Composition} & \textbf{Crystal Type} & \textbf{Space Group} & $\bm{T_C}$ (K) \\
\hline
smps-47828 & AlCo$_2$Fe & Tetragonal & P4/mmm (123) & 1200 \\
smps-792162 & Al$_2$Co$_3$Fe$_3$ & Orthorhombic & Pmmm (47) & 1090 \\
smps-70138 & Fe$_2$NiSi & Tetragonal & P4/mmm (123)  & 1078 \\
smps-1099935  & Fe$_2$RhTi  & Tetragonal & P4mm (99) &  1391 \\
smps-579308 & Co$_2$MnSi  & Tetragonal & P4/mmm (123) & 950  \\
smps-824124 & AlFe$_3$Pd$_4$ & Orthorhombic & Pmmm (47) & 832  \\
\hline
\end{tabularx}  
\end{table}

\begin{table}[htbp]
  \centering
  \caption{Néel temperature of the antiferromagnetic materials}
  \label{tab:AFM-TN}

\renewcommand{\arraystretch}{1.2}
\setlength{\tabcolsep}{3.2pt}
\rowcolors{2}{gray!10}{white}

\begin{tabularx}{\textwidth}{|L|L|L|L|C|}\hline
\rowcolor{gray!20}
\textbf{ID} & \textbf{Composition} & \textbf{Crystal Type} & \textbf{Space Group} & $\bm{T_N}$ (K) \\
\hline

smps-753026 & MgMn$_3$Pd$_4$ & Orthorhombic & Cmmm (65)  & 1059 \\
smps-960445 & MgMn$_3$Pd$_4$  & Monoclinic & P2/m (10)  & 1234 \\
smps-439167 & MgMn$_4$Pd$_{5}$ & Orthorhombic & Pmmm (47) & 1326 \\
smps-848080 & Mn$_2$Rh$_3$Ti  & Orthorhombic & Pmmm (47)  & 976  \\
\hline
\end{tabularx}  
\end{table}

\clearpage
\section*{Magnetic Anisotropy and Orbital Moment Analysis}
\begin{table}[htbp]
  \centering
  \caption{Site-resolved orbital magnetic moments (in $\mu_B$) of Fe$_2$ZnO$_3$ for $U=5.3$ eV, computed for magnetization along the crystallographic a- and c-directions.}
  \label{tab:Fe2ZnO3-orbmom}

\renewcommand{\arraystretch}{1.2}
\setlength{\tabcolsep}{3.2pt}
\rowcolors{2}{gray!10}{white}

\begin{tabularx}{\textwidth}{|l|C|C|}
\hline
\rowcolor{gray!20}
\textbf{Atom} & $\mathbf{m_a}$ ($\boldsymbol{\mu_B}$) & $\mathbf{m_c}$ ($\boldsymbol{\mu_B}$) \\
\hline

Fe & 0.026 & 0.096 \\
Zn & 0.0   & 0.0   \\
O  & 0.001 & -0.002 \\

\hline
\end{tabularx}  
\end{table}

\begin{table}[htbp]
  \centering
  \caption{Dependence of the magnetocrystalline anisotropy constant $K_1$ of Fe$_2$ZnO$_3$ on the Hubbard $U$ parameter applied to Fe 3d states.}
  \label{tab:Fe2ZnO3-K1-vs-U}

\renewcommand{\arraystretch}{1.2}
\setlength{\tabcolsep}{3.2pt}
\rowcolors{2}{gray!10}{white}

\begin{tabularx}{\textwidth}{|C|C|}
\hline
\rowcolor{gray!20}
\textbf{$U$ (eV)} & $\mathbf{K_1}$ (MJ/m$^3$) \\
\hline

3   & 4.718 \\
4   & 6.552 \\
5.3 & 6.848 \\
6   & 6.894 \\

\hline
\end{tabularx}  
\end{table}

\clearpage
\section*{Finite Temperature Stability}

\begin{figure}[H]
    \centering
    \includegraphics[scale=0.4]{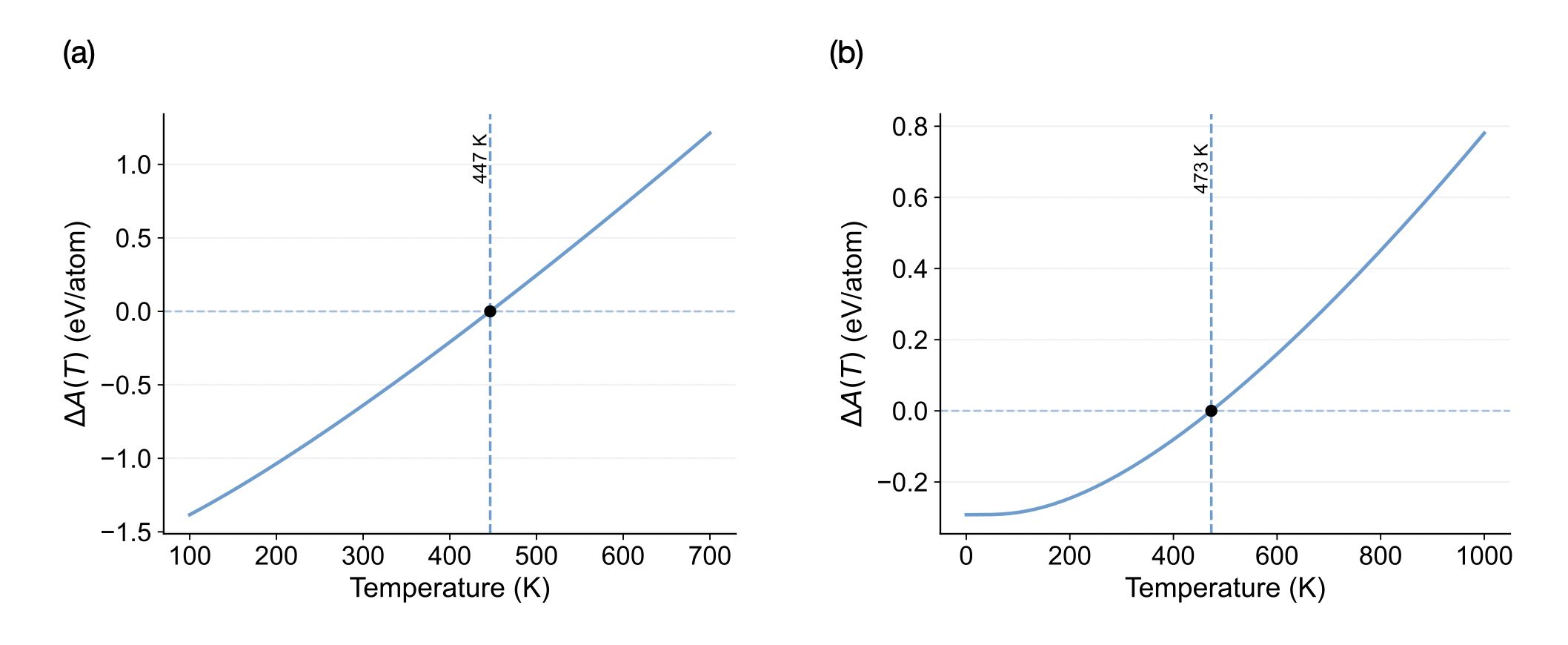}
    \caption{Temperature-dependent change in Helmholtz free energy of formation, $\Delta A(T)$ for (a) Fe$_2$ZnO$_3$ and (b) Fe$_2$NiSi.}
    \label{finite-T}
\end{figure}

\clearpage
\section*{Structural, Vibrational, and Electronic Properties of $\text{Fe}_2\text{NiSi}$}
\begin{figure}[H]
    \centering
    \includegraphics[scale=0.4]{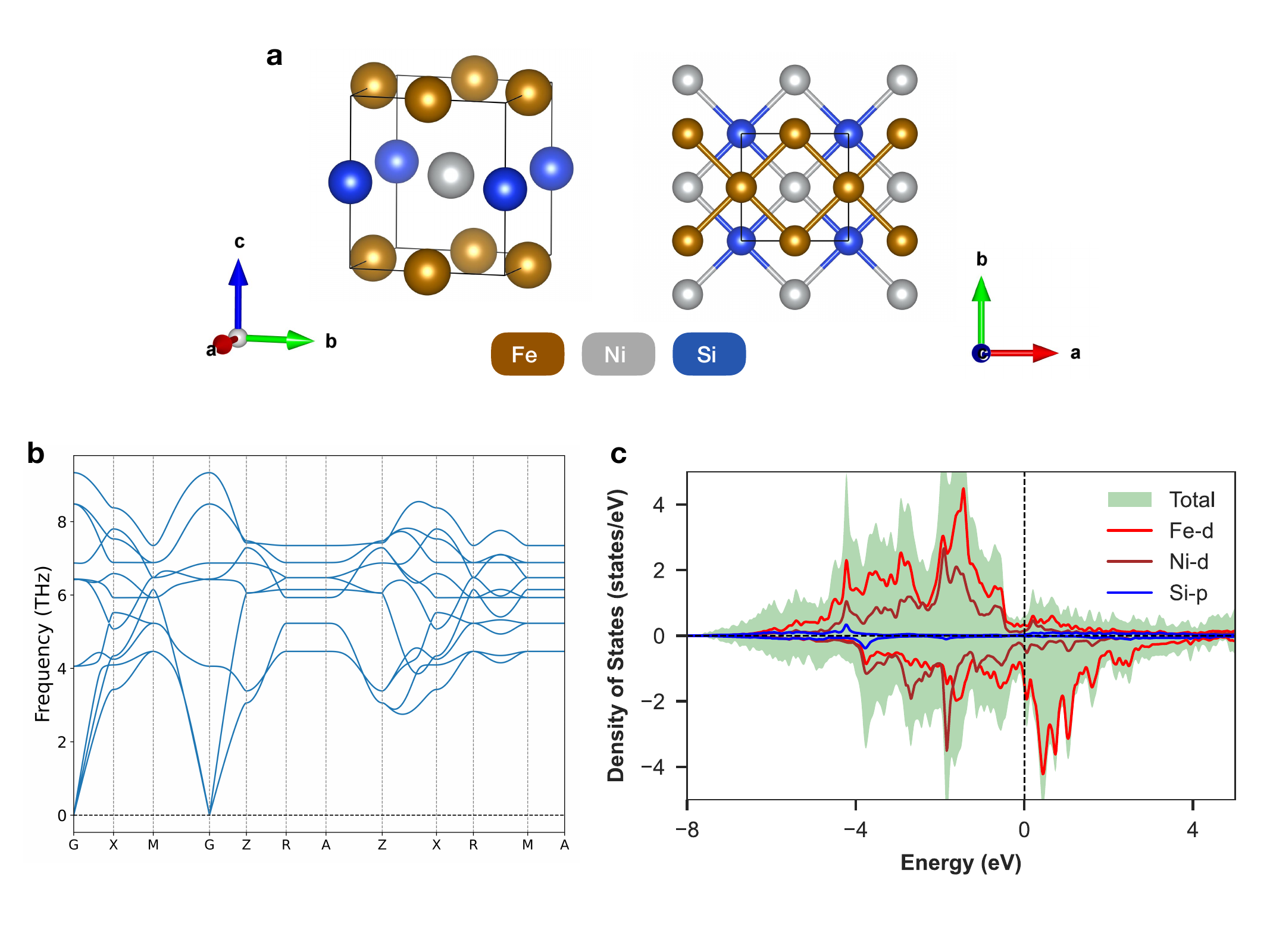}
   \caption{ {\bf Electronic, vibrational and magnetic properties of $\text{Fe}_2\text{NiSi}$.}
{\bf a} Side and top views of the $\text{Fe}_2\text{NiSi}$ crystal structure.
{\bf b} Phonon band structure along high-symmetry directions of the Brillouin zone, indicating dynamical stability of the system.
{\bf c} Total and atom-projected density of states (DOS). The shaded green region denotes the total DOS, while lines represent projections onto Fe-$3d$, Ni-$3d$, and Si-$3p$ orbitals. Positive and negative DOS correspond to the spin-up and spin-down channels, respectively, with the Fermi level ($E_F$) set to 0 eV. }
    \label{Fe2NiSi}
\end{figure}

\clearpage
\section*{Invertible Real‑Space Crystallographic Representation (IRCR).}
\begin{table}[ht]
  \centering
  \caption{Components of the invertible real‑space crystallographic representation (IRCR).}
  \label{tab:ircr-components}
\renewcommand{\arraystretch}{1.3}
\setlength{\tabcolsep}{4pt}
\rowcolors{2}{gray!10}{white}
\begin{tabularx}{\textwidth}{|C|C|L|}
\hline
\rowcolor{gray!20} \textbf{Component} & \textbf{Shape} & \textbf{Description} \\
\hline
$E$ (Element Matrix) & $94 \times 3$ & One-hot encoding of chemical elements (up to $Z=94$). \\
$L$ (Lattice Matrix) & $2 \times 3$ & Lattice constants ($a,b,c$) and angles ($\alpha,\beta,\gamma$). \\
$C$ (Coordinate Matrix) & $n_{\text{sites}} \times 3$ & Fractional atomic coordinates in the unit cell. \\
$O$ (Occupancy Matrix) & $n_{\text{sites}} \times 3$ & One-hot encoding of site occupancy. \\
$P$ (Property Matrix) & $6 \times 3$ & Elemental properties: atomic number ($Z$), electronegativity ($e$), period number ($p$), group number ($g$), valence electron number ($n_v$), and atomic fraction ($s$). \\
\hline
\end{tabularx}
\end{table}

\clearpage
\section*{Analysis of Training Set Data}
\begin{figure}[H]
    \centering
    \includegraphics[scale=0.35]{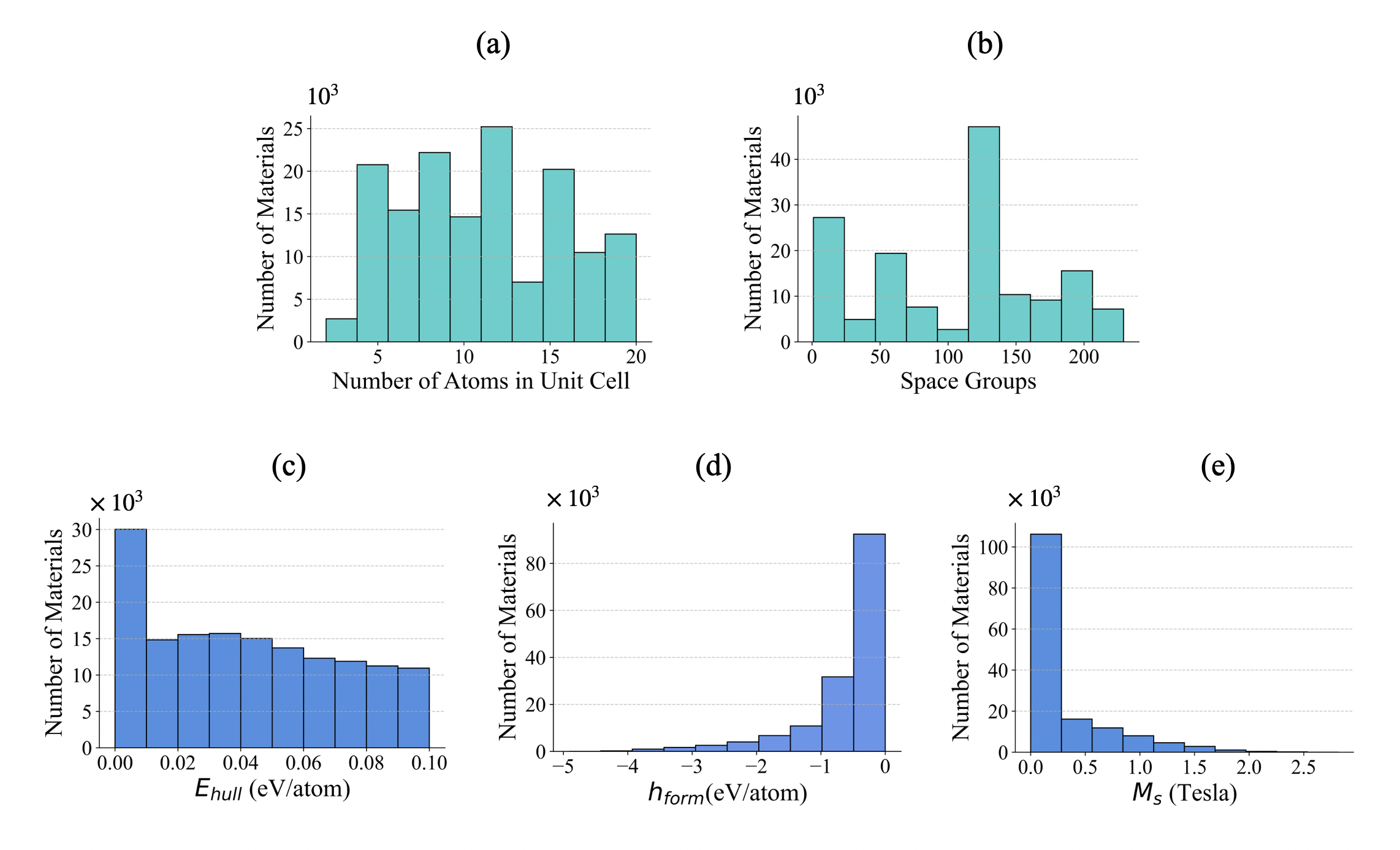}
   \caption{{\bf Analysis of training set data } Distribution of (a) number of atoms in unit cell, (b) space groups, (c) convex hull, (d) formation energy and (e) saturation magnetization of materials in the Alex-1 dataset.}
    \label{training-data}
\end{figure}

\clearpage 
\section*{Training Dynamics of the Diffusion Model and Prediction Accuracy of Property Predictors}
\begin{figure}[H]
    \centering
    \includegraphics[scale=0.4]{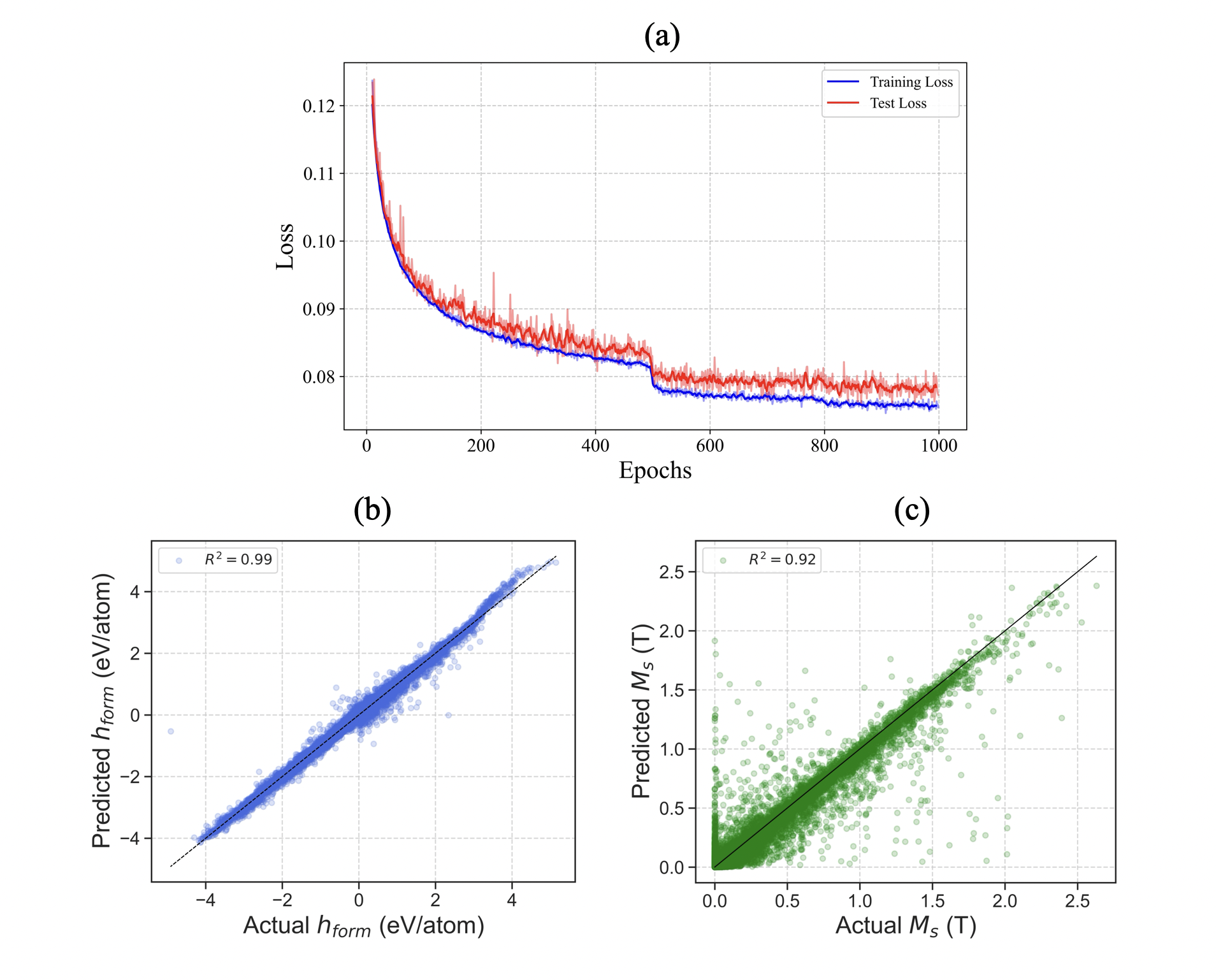}
    \caption{(a) Learning curve of the diffusion model showing training loss (blue) and test loss (red) over 1000 epochs, indicating stable convergence. (b) Parity plot of predicted versus actual $h_{form}$ in test set, achieving an $R^2=0.99$. (c) Parity plot of predicted versus actual $M_s$ in test set, achieving an $R^2=0.92$.}
    \label{parity-plot}
\end{figure}

\clearpage